\documentclass[fleqn,usenatbib]{mnras}
\usepackage[utf8]{inputenc}

\usepackage{newtxtext,newtxmath}
\usepackage[T1]{fontenc}

\usepackage{graphicx}
\usepackage{amsmath}
\usepackage{natbib}
\usepackage{bm}
\usepackage{xcolor}
\usepackage{soul} % Gives /st to strikethrough

\DeclareMathAlphabet{\mathsfit}{T1}{\sfdefault}{\mddefault}{\sldefault}
\SetMathAlphabet{\mathsfit}{bold}{T1}{\sfdefault}{\bfdefault}{\sldefault}

\newcommand*{\tensor}[1]{\bm{\mathsfit{#1}}}
\newcommand*{\myvec}[1]{\bm{#1}}

\newcommand{\Mpc}{~\mathrm{Mpc}}

\newcommand{\Mo}{~\mathrm{M}_\odot}

\newcommand{\kmsMpc}{~\mathrm{km}\,\mathrm{s}^{-1}\,\mathrm{Mpc}^{-1}}

\usepackage{xspace}
\newcommand{\gadget}{\textsc{gadget}\xspace}
\newcommand{\gadgettwo}{\textsc{gadget-2}\xspace}
\newcommand{\gadgetfour}{\textsc{gadget-4}\xspace}
\newcommand{\gasoline}{\textsc{gasoline}\xspace}
\newcommand{\gasolinetwo}{\textsc{gasoline2}\xspace}
\newcommand{\changa}{\textsc{ChaNGa}\xspace}
\newcommand{\ramses}{\textsc{ramses}\xspace}
\newcommand{\enzo}{\textsc{enzo}\xspace}
\newcommand{\pkdgrav}{\textsc{pkdgrav}\xspace}
\newcommand{\pkdgravthree}{\textsc{pkdgrav3}\xspace}
\newcommand{\arepo}{\textsc{arepo}\xspace}
\newcommand{\gizmo}{\textsc{gizmo}\xspace}
\newcommand{\swift}{\textsc{swift}\xspace}
\newcommand{\sphantom}{\textsc{phantom}\xspace}
\newcommand{\shadowfax}{\textsc{shadowfax}\xspace}
\newcommand{\athena}{\textsc{athena}\xspace}
\newcommand{\sphgal}{\textsc{SPHGal}\xspace}
\newcommand{\swiftcode}{\textsc{swift}\xspace}
\newcommand{\sphenix}{\textsc{sphenix}\xspace}
\newcommand{\ahf}{\textsc{ahf}\xspace}

\title[Mesh-free hydrodynamics in \textsc{pkdgrav3}]{Mesh-free hydrodynamics in \textsc{pkdgrav3} for galaxy formation simulations}

\author[I. Alonso Asensio \textit{et al.}]{
Isaac Alonso Asensio$^{1,2}$\thanks{E-mail: isaacaa@iac.es},
Claudio Dalla Vecchia$^{1,2}$,
Douglas Potter$^3$ and
Joachim Stadel$^3$
\\
$^1$Instituto de Astrof\'isica de Canarias, C/V\'ia L\'actea s/n, E-38205 La Laguna, Tenerife, Spain  \\
$^2$Departamento de Astrof\'isica, Universidad de La Laguna, Av.~Astrof\'isico Francisco S\'anchez s/n, E-38206 La Laguna, Tenerife, Spain \\
$^3$Institute for Computational Science, University of Zurich, Winterthurerstrasse 190, 8057 Zurich, Switzerland
}

\begin{document}

\maketitle

\begin{abstract}
We extend the state-of-the-art N-body code \textsc{pkdgrav3} with the inclusion of mesh-free gas hydrodynamics for cosmological simulations.
Two new hydrodynamic solvers have been implemented, the \textit{mesh-less finite volume} and \textit{mesh-less finite mass} methods.
The solvers manifestly conserve mass, momentum and energy, and have been validated with a wide range of standard test simulations, including cosmological simulations.
We also describe improvements to \textsc{pkdgrav3} that have been implemented for performing hydrodynamic simulations.
These changes have been made with efficiency and modularity in mind, and provide a solid base for the implementation of the required modules for galaxy formation and evolution physics and future porting to GPUs.
The code is released in a public repository, together with the documentation and all the test simulations presented in this work.
\end{abstract}


%%%%%%%%%%
%%%%%%%%%%
%%%%%%%%%%

\section{Introduction}

Cosmological hydrodynamic simulations are key in our understanding of the formation and evolution of galaxies.
However, performing these simulations is a daunting task due to the complexity of the numerical codes needed, the models and algorithms implemented in them and the very large computing time required for achieving the expected resolution and/or statistics.

All modern cosmological simulations require the use of High Performance Computing facilities hosting powerful supercomputers.
Researchers must carefully develop codes that use these systems efficiently.
The development, debugging and testing of codes can take a substantial amount of time, resulting in only a handful of codes that can perform galaxy evolution studies.
As a result, the possibility to reproduce or compare different methods, or even different implementations of the same method, is rather limited.
Furthermore, not all the simulation codes are fully open source, and therefore examining the details of their methods may be impossible.

Despite of this, during recent years there has been an increase in the number of cosmological hydrodynamic simulations produced by leading teams:
the Illustris \citep{Vogelsberger2014} and IllustrisTNG \citep{Pillepich2018} simulations, performed with \arepo \citep{Springel2010};
the HorizonAGN simulations \citep{Dubois2014}, performed with \ramses \citep{Teyssier2002};
the EAGLE \citep{Schaye2015}, C-EAGLE \citep{Barnes2017}, and BAHAMAS \citep{McCarthy2017} simulations, performed with modified versions of \gadget \citep{Springel2005b};
and the MUFASA/SIMBA simulations \citep{Dave2016,Dave2019}, performed with \gizmo \citep{Hopkins2015}, just to cite a few.
Different codes and methods are employed for evolving the baryonic component (e.g., gas dynamics, cooling, star formation, feedback, etc.), yet they provided a somewhat consistent view of the formation of galaxies \citep[][and references therein]{Somerville2015,Vogelsberger2019}.

However, there are still uncertainties on the physical processes shaping galaxies and their numerical modelling in cosmological hydrodynamic simulations.
Due to the limited resolution of current simulations, most physical processes must be approximated by the so-called sub-grid modelling that reproduces the macroscopic behaviour of unresolved physics.
The most notable example is the feedback from supernovae and active galactic nuclei, that can drastically change the evolution of galaxies \citep{Schaye2010, Dubois2013, Crain2015, Rosito2020}, but for which there is no consensus model.
Therefore, there is the need to develop new methods and codes to compare their output among them and against forthcoming or existing observations.

With this work, we aim to provide the community a new, efficient and modular code to perform galaxy formation and evolution studies.
Rather than developing the code from scratch, we have adopted and extended the state-of-the-art, GPU-accelerated N-body code \pkdgravthree \citep{Potter2017}.
The choice comes from the proved efficiency of \pkdgravthree in running simulations with trillions of particles, its low-level structure and support for Graphics Processing Unit (GPU) acceleration.

As mentioned above, there are several schemes and algorithms for simulating gas dynamics in cosmological hydrodynamic simulations.
In general, they can be subdivided as either mesh-based or particle-based.
The former includes schemes that partition the computational domain in volume elements (cells), whose hydrodynamic state is evolved over time.
Cells are typically arranged in a Cartesian mesh that can be adaptively subdivided in smaller cells to increase the resolution wherever needed.
Cell-based volume partitions can be easily structured in memory allowing for very efficient codes. However, they are not Galilean invariant.
Among the codes using Cartesian meshes and designed for astrophysical hydrodynamic simulations are  \ramses \citep{Teyssier2002}, \athena \citep{Stone2008} and \enzo \citep{Bryan2014}.

On the other hand, particle-based schemes are Lagrangian and Galilean invariant.
Instead of volume, mass in the computational domain is partitioned into finite mass elements (particles).
These schemes integrate the equations of motion of the set of particles, which also carry thermodynamic information.
The most used particle-based method is Smoothed Particle Hydrodynamics (SPH) \citep{Lucy77, Gingold77}.
In this method, the fluid density is directly related with the number density of particles, thus the resolved spatial scale varies adaptively with the local density, with denser regions having higher spatial resolution \citep[see][for a review]{Rosswog2009,Price2012}.
In its original formulation, SPH conserves mass and linear and angular momentum by construction. It is also non-diffusive and inviscid, and requires the modelling of viscosity to capture shocks and  special treatment of contact discontinuities \citep{Agertz2007,Cullen2010}.
Among the SPH codes used in astrophysics are \sphantom \citep{Price2018}, \gadget \citep{Springel2005b,Springel2021}, \gasoline \citep{Wadsley2004}, \gasolinetwo \citep{Wadsley2017}, \changa \citep{Menon2015}, \sphgal \citep{Hu2014} and \swiftcode \citep{Schaller2016} in its SPH formulation, \sphenix \citep{Borrow2022}.

There are, however, hydrodynamic methods that do not fit exactly in any of the above categories:
unstructured, moving mesh codes \citep[e.g.~\arepo and \shadowfax,][]{Springel2010,Vandenbroucke2016,Weinberger2020}, where the equations of hydrodynamics are solved on a Voronoi or Delaunay mesh that is built dynamically at run time.
Their quasi-Lagrangian formulation allows for adaptive resolution and Galilean invariance, at the cost of having to build and store an unstructured mesh in memory.
Other examples are the Meshless Finite Mass (MFM) and Meshless Finite Volume (MFV) methods implemented in \gizmo \citep{Hopkins2015}, and based on the theoretical work of \citet{Lanson2008a, Lanson2008b} and the implementation of \citet{Gaburov2011}.
In this case, the fluid is discretised in particles or tracers, and the fluid equations in each point in space are solved over a local mesh defined by the nearest neighbours.
This provides the solution of the hydrodynamics equations with accuracy similar to that of mesh-based methods and adaptive resolution as in particle-based methods, without the need of storing the mesh in memory.
However, solving the Riemann problem for each particle-neighbour interaction is required, making  both methods computationally expensive.
In this work we have implemented both MFV and MFM.
As they are particle-based, they couple nicely with the already working \pkdgravthree gravity implementation.

The article is organized as follows.
The optimizations of the original version of \pkdgravthree and the implementation of the hydrodynamic solvers are described in section~\ref{sec:Methods}.
Numerical hydrodynamic tests are shown in section~\ref{sec:Results}.
Summary and final remarks are given in section~\ref{sec:Conclusions}.

%%%%%%%%%%

\section{Code development} \label{sec:Methods}

\subsection{Code description} \label{sec:Code_description}

\pkdgrav was first described in \citet{Stadel2001}.
Since its release, it has been employed in several numerical projects:
it has been compared directly with other N-body codes \citep{Power2003, Diemand2004};
it has been adapted to perform planetesimal dynamics simulations \citep{Richardson2000} and thenceforth widely used in this field \citep[e.g.,][]{Leinhardt2000, Leinhardt2009,Nesvorny2010};
it has been the framework for the development of the SPH code \gasoline \citep{Wadsley2004} and its last version \gasolinetwo \citep{Wadsley2017}, which have been employed in numerous studies of galaxy formation and evolution \citep[e.g., the NIHAO project, ][]{Wang2015}.

\pkdgrav pure N-body version was accelerated by porting computationally demanding parts of the code to GPU, and renamed as \pkdgravthree \citep{Potter2017}.
The code is written in \texttt{C}/\texttt{C++}, and the GPU acceleration in \texttt{CUDA} (although it can be run without GPUs).
The configuration and compilation of the code is managed through \texttt{cmake}/\texttt{make}, and the source code is under version control with \texttt{git}.
The parallelization is handled by \texttt{MPI} for inter-node communication, and \texttt{pthreads} for fully exploiting the shared memory within single nodes.
From the developer's perspective this is completely transparent (see section~\ref{sec:paralellization}).
Most of the behaviour of the code is defined in the parameters file used at run time, although some general options are set at compile time.
\pkdgravthree provides an on-the-fly friends-of-friends (FOF) halo finder, light cone output and matter power spectrum calculator, as well as a cosmological initial conditions generator.
This version has also been compared to other N-body codes in \citet{Schneider2016} and \citet{Garrison2019}.
One of the largest simulations of the large-scale structure of the Universe to date has been performed with \pkdgravthree.
This simulation is being used by the Euclid Collaboration \citep{Euclid2019}.

The version of \pkdgravthree developed in this work is released under the GNU General Public License (version 3, GPLv3).\footnote{\url{https://www.gnu.org/licenses/gpl-3.0.en.html}}
To ease the adoption of \pkdgravthree, and as an exercise of open science, source code, parameters files and initial conditions used to generate the figures in this paper are provided in public repositories.\footnote{They can be accessed through the project's webpage: \url{https://research.iac.es/proyecto/PKDGRAV3}}

When implementing the hydrodynamic solver into \pkdgravthree, we needed to add a new type of particle (gas particle) and modify part of the original code to gain in performance when treating baryons, such as the neighbours finding algorithm.
Major changes to the code with detailed technical improvements are presented in section~\ref{sec:Code}.
The reader less interested in these technicalities may skip to section~\ref{sec:Hydrodynamics}, where the implementation of the hydrodynamic solver is described.

\subsection{Technical improvements} \label{sec:Code}

\subsubsection{Memory management} \label{sec:memory}

In contrast with broadly used particle-based hydrodynamic codes
we use a single, dynamically created particle structure, independent of the type of particle (dark matter, gas, star, black hole or whatever user-defined type).
Having a unique particle structure has one main advantage: memory locality.
As there is no pointer to external arrays in the particle structure, all particle information is stored contiguously in memory.
This leads to higher performance when fetching data from the RAM to the CPU cache, as less scattered memory accesses are required for reading/writing the data of a particle.
Furthermore, particle handling is easier and more transparent, allowing to reorder the particle array without requiring any bookkeeping.
Performance can degrade if the memory occupied by one particle is too large and cache misses occur when accessing different particle variables.
However, the worst case scenario would be similar to having the particle data scattered across different locations in memory (e.g., when using more than one array to store a particle type), as is often done in other codes.

The particle structure size is defined by the particle type occupying the largest size in memory.
This means that not all memory allocated for particles is used.
We have mitigated this using the C \texttt{union} construct.
Values common to all particles (e.g., coordinates) are allocated in the first part of the particle's memory block, whilst the union that contains variables specific to the particle type are allocated in the reminder memory.

\begin{figure}
   \centering
   \includegraphics[width=0.95\linewidth]{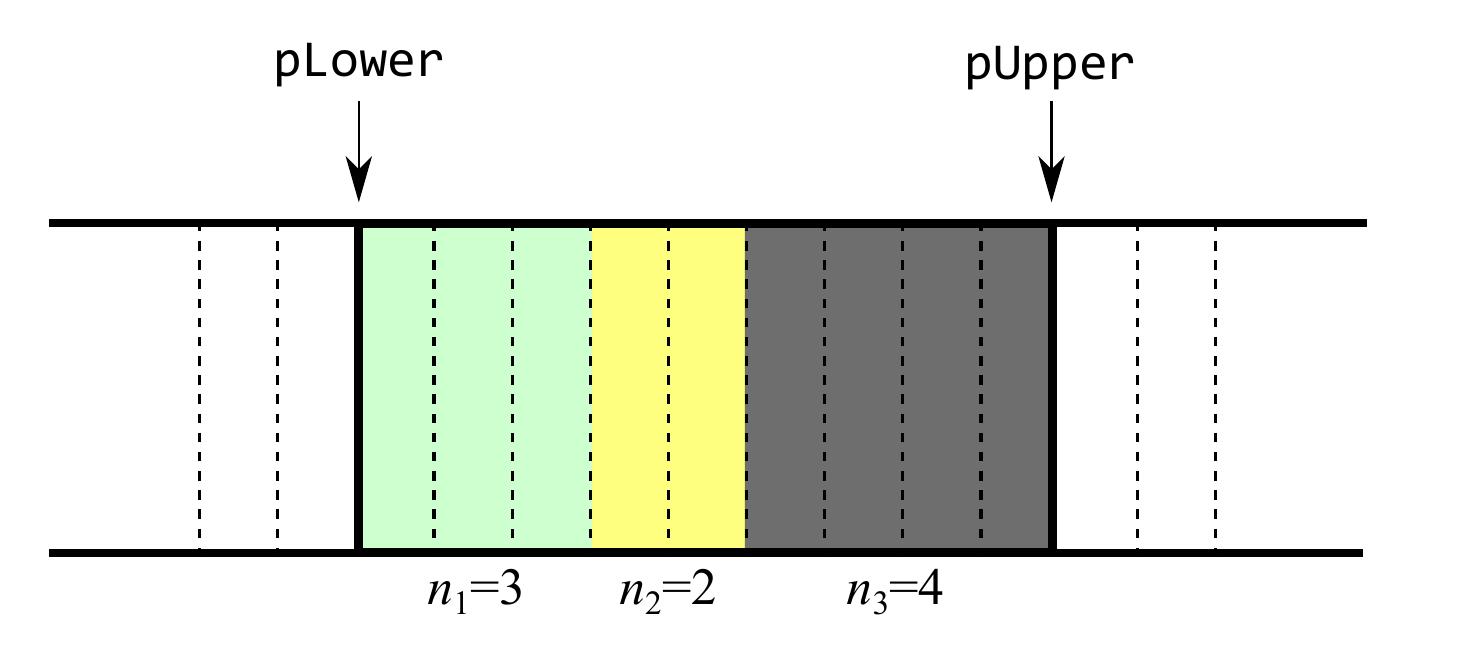}
   \caption{Tree node-based memory layout of the particle array. All the particles belonging to a tree node are stored contiguously in memory.
   The node stores the pointers to the beginning and end of its memory domain (\texttt{pLower} and \texttt{pUpper}, respectively), together with the number of particles of each type it contains ($n_1$, $n_2$ and $n_3$ in this example).
   This allows for fast particle loops without explicit type checking, reducing branch misprediction and data accesses.}
   \label{fig:node_memory_layout}
\end{figure}

As stated above, particles can be seamlessly reordered in memory to improve performance when needed.
This advantage is exploited in the code whenever possible.
For example, after building the tree structure, particles belonging to a given tree node are arranged contiguously in memory, such that each node has a \texttt{pLower} and \texttt{pUpper} pointer to its first and last particle, respectively.
In the case that the tree node does not contain any further subdivision, it is termed a \textit{leaf} node, and contains a maximum amount of particles: the \textit{bucket} size.
Inside the leaf nodes, we further sort particles by type, as shown in figure~\ref{fig:node_memory_layout}.
This allows for a fast loop over particles in a spatially ordered way, without any explicit check of the particle type, thus substantially reducing accesses to particle data and branch misprediction.

As another example, the current implementation of the on-the-fly FOF halo finder sorts the particle array according to the group particles belong to.
This way, computing groups' global properties is faster, as the particle data can be efficiently accessed and cached.
The only caveat of the frequent reordering of the particle array is that the developer must be aware of the current order of particles.
As can be guessed from the above examples, doing a FOF search will render the tree unusable, and tree ordering must be reestablished after each FOF task.

\subsubsection{Parallelisation strategy} \label{sec:paralellization}

The code inherits the parallelisation strategy of \pkdgravthree \citep{Stadel2001,Potter2017}, and no major modifications have been carried out.
For the sake of completeness, we include here a brief description of its current status.

\pkdgravthree is organized in four abstraction layers.
These allow for a clean separation between communication and the actual computation, such that the developer does not need to know the details of, or code, the communication patterns.
This organization also provides high adaptability to system architectures, cluster topologies and scientific problems.
We briefly describe the four abstraction layers in the code. These are:
\begin{enumerate}
   \item The Machine Dependent Layer (MDL), the deepest layer in charge of communication, task managing and scheduling.
      This layer caches frequently accessed data from other processes to reduce message passing.
      It can handle \texttt{pthreads} for shared memory and \texttt{MPI} for distributed memory systems, or a combination of them.
      It is also in charge of offloading the work to the GPUs, if available.
   \item The Master Layer (MSR), executed on one process serially.
      It defines the workflow of the program, from reading the initial parameters to dispatching computation to the other processes via the next layer.
   \item The Processor Set Tree (PST), which organizes the processes in a binary tree, the master process being its root node.
      When some computation is dispatched to the PST, each node passes the information to its branches recursively, and then starts its own part of the computation, which is handled by the last layer.
   \item The shallowest layer, PKD, is the only layer that is allowed to modify particles/tree data.
      It is executed on all cores and encloses all the physics modules.
      This layer is quasi-serial code that operates on data local to the core. Most algorithms (e.g., force calculations) require data that is “remote” and this layer uses MDL to fetch such data on demand. This functionality operates very similar to PGAS (partitioned global address space) where data is exposed as a virtual vector and remote elements are accessed by index.
\end{enumerate}
Furthermore, specific functions can operate between the PST and PKD layers.
One example is the neighbours finding algorithm (section~\ref{sec:neighbours}).
This function is called within the PST layer, and provides data to the PKD layer.
In practise, it queries the MDL to request the data from remote processes such that the PKD layer can be blind to their origin.

\subsubsection{Neighbours finding} \label{sec:neighbours}

The hydrodynamic scheme requires an efficient implementation of a neighbours finding algorithm.
Indeed, most of the calculations can be found within loops over the closest neighbouring particles, i.e., those that lie inside the compact support kernel of a given particle (see section~\ref{sec:Hydrodynamics}).
The hydrodynamic scheme can be separated in four different loops (or smoothing operators, following the nomenclature of SPH) over neighbouring particles:
(1) smoothing length and number density computation;
(2) gradients estimation and limiter;
(3) temporal/spatial extrapolation and Riemann solver;
(4) time step computation.
These take most of the computational time in our implementation of the hydrodynamic scheme, thus it is key to improve their performance, and the first step towards that is an efficient neighbours finding procedure.

We have experimented with two different algorithms.
The first one is a loop over particles and, for each particle, a walk of the tree searching for the closest neighbours.
This is what is generally done in tree codes such as \gadget
or \gasolinetwo,
and has the advantage of being straightforward to implement.
Moreover, this may be the only viable way to implement smoothing operators in codes that explicitly split the particles loop between local and remote neighbours finding.

We have found that this approach leads to important bottlenecks in the smoothing operators due to excessive communication and tree walks.
In some cases, the neighbours finding can even be more time consuming than solving the Riemann problem, which is far from ideal.
Moreover, we implemented the hydrodynamics in \pkdgravthree having in mind the eventual porting to GPUs, and doing a tree walk per particle is not a task suitable for GPU acceleration.

The second approach is to group particles together (in buckets), and perform a single neighbours search for all the particles in the bucket.
This has the advantage that less tree walks are performed, and, in general, less communication is needed.
We have decided to implement this approach, as it is more consistent with the gravity tree walk algorithm of \pkdgravthree, and is also used in codes like \swift \citep{Schaller2018} and \changa \citep{Menon2015}.

Our implementation is as follows:
the pointers to the particles that can potentially interact with any of the particles in a bucket are gathered.
For each particle in the bucket a buffer is filled with the required information of the interacting neighbouring particles using a structure of arrays (SoA).
This step can be omitted, allowing direct access the particle data.
Finally, a loop is performed over these buffers to compute the hydrodynamic interaction, and the result is added to the particle and to the corresponding neighbours.

If the input buffer is used with SoA ordering, the smoothing operator can be accelerated with CPU vector instructions (e.g., SSE, AVX).
The SoA ordering incurs in certain slowdown of the code, as it involves more memory movement in an already memory bounded region of the code.
However, as it opens the door for vectorization, the slow-down can be completely mitigated.
Intel\texttrademark\ compilers can successfully (auto)vectorize the whole spatial/temporal extrapolation and Riemann solver loop \citep[see][for a vectorization guide]{Intel2012},\footnote{However, GNU or LLVM-based compilers can not yet autovectorise the loop. We are working on an explicit SIMD intrinsic approach to avoid being compiler-locked.} achieving an overall 40 percent speedup of the flux computation.
If vectorization is disabled, a 20 percent slowdown is observed due to the SoA ordering overheads.\footnote{This timing were obtained setting a $N=2\times64^3$ cosmological box at $z=49$ in an Intel\texttrademark\ Xeon\texttrademark\ Processor E5-2670 using 8 threads. We note that this CPU is rather old and only supports AVX2 (SIMD register width of 256 bits), newer CPUs with AVX-512 support show larger speedups.}

For the rest of the hydrodynamics loops, as they are not as strongly compute intensive we have opted to omit the SoA ordering.
In this case, using the new neighbour search yield a speedup that ranges between 20 and 50 percent by reducing the number of tree walks and data accesses.
In addition, the structure ordering will facilitate an eventual port of the hydrodynamic loops to GPUs.
This will be part of future work.

\subsection{Hydrodynamics solver} \label{sec:Hydrodynamics}

The development of the hydrodynamic solver follows the theoretical work of \citet{Lanson2008a, Lanson2008b} and the implementations of \cite{Gaburov2011} and \cite{Hopkins2015}.
The solver was chosen due to its superior performance in treating shocks and contact discontinuities when compared to SPH \citep{Hopkins2015}.
This comes at the price of increased computational complexity and computational cost.
However, as mentioned in section~\ref{sec:neighbours}, we applied several technical solutions for both increasing the performance of the solver and prepare it for the porting to GPU.
We describe in the following sections the mathematical derivation of the numerical scheme and all the methods that form the hydrodynamic solver.

\subsubsection{Weak solution of the fluid equations}

The weak solution of the fluid equations, which is the base of the hydrodynamic scheme, is derived first.
The fluid equations in a moving frame of reference with velocity $\myvec{v}_\text{frame}$ are
\begin{equation}
   \frac{\partial \myvec{U}}{\partial t} + \nabla \cdot \left( \myvec{F} + \myvec{v}_\text{frame} \otimes \myvec{U} \right) = \myvec{S}.
   \label{eq:cons}
\end{equation}
The conservative variables, their fluxes and the source terms are, respectively,
\begin{equation}
   \myvec{U} = \begin{pmatrix}
      \rho \\
      \rho \myvec{v} \\
      \rho e
   \end{pmatrix},
   \hspace{0.5cm}
   \myvec{F} = \begin{pmatrix}
      \rho \myvec{v} \\
      \rho \myvec{v} \otimes \myvec{v} + p \tensor{I} \\
      \left( \rho e + p \right) \myvec{v} \\
   \end{pmatrix}
   \hspace{0.5cm}\mathrm{and}\hspace{0.5cm}
   \myvec{S} = \begin{pmatrix}
      S_\mathrm{M} \\
      \bm S_\mathbf{P} \\
      S_\mathrm{E}
   \end{pmatrix},
   \label{eq:fluid_variables}
\end{equation}
where $S_\mathrm{M}$, $\bm S_\mathbf{P}$ and $S_\mathrm{E}$ are any source terms for mass, momentum and energy.
Total mass, $M=\int_V \rho \mathrm{d}V$, linear momentum, $\myvec{P}=\int_V \rho \myvec{v} \mathrm{d}V$, and total energy, $E=\int_V \rho e \mathrm{d}V$, are conserved in the absence of source terms. In the above equations, $\rho$, $\myvec{v}$, $p$ and $e$ are the fluid density, velocity, pressure and total energy per unit mass. The identity matrix, $\tensor{I}$, and the outer vector product, $\otimes$, are
\begin{equation}
    \tensor{I} = \begin{pmatrix}
    1 & 0 & 0 \\
    0 & 1 & 0 \\
    0 & 0 & 1 \\
    \end{pmatrix}\quad\mathrm{and}\quad
    \myvec{v} \otimes \myvec{v} = \begin{pmatrix}
    v_x v_x & v_x v_y & v_x v_z \\
    v_y v_x & v_y v_y & v_y v_z \\
    v_z v_x & v_z v_y & v_z v_z \\
    \end{pmatrix}.
\end{equation}

The fluid equations, as written above, imply that $\myvec{U}$ and $\myvec{F}$ are differentiable.
However, due to the non-linearity of the system of equations, discontinuities may develop in the domain.
To relax the condition of differentiability, a weak form of the equations can be derived.
Following, for example, \cite{Vila99} and \cite{Lanson2008a},
\begin{equation} \label{eq:weak_solution}
   \int_{\mathbb{R}^\nu \times \mathbb{R}^+} \left[ \myvec{U} \frac{\mathrm{d}\varphi}{\mathrm{d}t} + \myvec{F}\nabla \varphi - \myvec{S} \varphi \right] \mathrm{d}^\nu\!\myvec{x}\mathrm{d}t=0\,,
\end{equation}
where $\varphi (\myvec{x}, t)$ is a differentiable test function with compact support in $\nu$-dimensional space and time,  $\mathbb{R}^\nu \times \mathbb{R}^+$.
The advective derivative of a function $f(\myvec{x}, t)$ is defined as $\mathrm{d}f/\mathrm{d}t = \partial f/\partial t + \myvec{v}_\text{frame} \nabla f$.

For the further development of the numerical scheme, three basic ingredients are needed: a discretization of the spatial integral, an estimate of the $\nabla \varphi$ term, and a discretization of the time differential. These are detailed in the following sections.

\subsubsection{Partition of unity}  \label{sec:partition_unity}

In principle, the spatial integral in equation~(\ref{eq:weak_solution}) could be arbitrarly discretized.
The discretization could be based on a grid (or mesh) of tracer points defining volume elements, whose geometrical faces can be determined, and the Divergence Theorem applied to the $\nabla \varphi$ terms.
This would lead to a \textit{mesh-based} method.
If the tracers were allowed to move with the fluid, the mesh also would, leading to a \textit{moving-mesh} method.

In the present work, however, we focus on implementing a \textit{meshless} method, in which neither a geometrical volume nor faces are defined for the tracers.
This method has the advantage that no connectivity information for the whole mesh needs to be computed and/or stored, which is time and memory consuming.
On the other hand, as geometrical faces or volumes cannot be defined, the Divergence Theorem cannot be applied directly, and other means to treat the $\nabla \varphi$ term are required.

Although the geometry cannot be precisely specified, if a set of $N$ tracers is known, with coordinates ${\myvec{x}_i \in \mathbb{R}^\nu, \forall i\in[1,N]}$, then, for any position, the volume partition can be defined as
\begin{equation}
   \psi_i (\myvec{x}) = \omega (\myvec{x}) W (\myvec{x}-\myvec{x}_i, h(\myvec{x}))\,,
\end{equation}
where
\begin{equation}
    \omega^{-1} (\myvec{x}) = \sum_{j=1}^N W (\myvec{x}-\myvec{x}_j, h(\myvec{x}))\,,
\end{equation}
such that the differential volume $\mathrm{d}^{\nu}\myvec{x}$ at position $\myvec{x}$ is the sum of the contribution $\psi_i(\myvec{x})$ of each tracer.
The kernel function, $W$, dictates how the contribution to $\mathrm{d}^{\nu}\myvec{x}$ decreases with distance.
In order to preserve isotropy and locality, the kernel is required to be a function of distance, $W(r_{ij}, h)$, where $r_{ij} = |\myvec{x}_i - \myvec{x}_j|$, and vanish beyond the compact support radius, $W(r, h)=0$, $\forall r>h$.
Furthermore, it is convenient to chose a continuous, derivable function.
The compact radius, $h$, is related to the resolution scale or \textit{smoothing length} by a factor which depends on the kernel being used \citep{Dehnen2012}.
In our case we use the cubic spline kernel, thus the smoothing length is $\approx 0.55$ times the compact support radius.
Although not explicitly needed (because $\psi_i(\myvec{x})$ is normalized by $\omega(\myvec{x})$), we require that $\int W(x, h)\,\mathrm{d}x = 1$ so that $\omega^{-1}(\myvec{x})$ becomes a measure of the local number density of tracers.
Consequently, $\omega(\myvec{x}_i)$ is an appropriate estimate of the particle's volume, $V_i$,
and the gas density of the $i$-th tracer is $\rho_i=m_i \omega^{-1}(\myvec{x}_i)$.

It is trivial to show that $\sum_i \psi_i(\myvec{x}) = 1$ for any $\myvec{x}$.
This property can be used to discretize the volume integral of any function, $f(\myvec{x})$:
\begin{align}
      \int_{\mathbb{R}^\nu} f(\myvec{x})\mathrm{d}^\nu\!\myvec{x} &= \sum_{i=1}^N \int f (\myvec{x}) \psi_i (\myvec{x})\mathrm{d}^\nu\!\myvec{x} \approx \nonumber \\
      &\approx \sum_{i=1}^N f_i \int \psi_i(\myvec{x})\mathrm{d}^\nu\!\myvec{x} + \mathcal{O}(h^2) = \sum_{i=1}^N f_i V_i + \mathcal{O}(h^2)\,, \label{eq:partition_volume}
\end{align}
where we have adopted the notation $g_i \equiv g(\myvec{x}_i)$ to refer to the value of a field at the position of the $i$-th tracer.\footnote{Notice that when referring to $\psi_i(\myvec{x})$, we can not drop the spatial dependence.}

In the above derivation, we have neglected any variation of the compact support, $h(\myvec{x})$.
However, in most application that require spatial adaptive resolution, the smoothing length  will continuously adapt to the local distribution of tracers (see section~\ref{sec:smoothing_computation})

Equation~\ref{eq:partition_volume} can be applied directly to the weak solution (equation~\ref{eq:weak_solution}) obtaining
\begin{equation} \label{eq:weak_solution_partition}
   \int_{\mathbb{R}^+} \sum_i V_i \left[ \myvec{U}_i \frac{\mathrm{d}\varphi_i}{\mathrm{d}t} + \myvec{F}_i(\nabla \varphi)_i - \myvec{S}_i \varphi_i \right] \mathrm{d}t= 0\,.
\end{equation}

We still need to define the gradient operator, which is the next step in the derivation of the equations of hydrodynamics. This is described in the following section.

\subsubsection{Gradient estimator} \label{sec:gradient_estimator}

In mesh-based schemes, where grid points are arranged in a Cartesian grid, there are well established methods to compute spatial gradients \citep{LeVeque1992}.
These schemes can be of very high order, which make them well suited to study phenomena involving small spatial scales and low diffusivity.
When the mesh is unstructured, it is still plausible to obtain good gradient estimators, although in general of lower order \citep{Barth89}.
For example, the gradient estimator used in \arepo \citep{Springel2010} for a Voronoi mesh, is second order accurate.

In SPH, a straightforward gradient definition is \citep[see e.g,][]{Rosswog2009,Price2012}
\begin{equation} \label{eq:SPH_grad}
   (\nabla f)_i = \sum_j \nabla[f_j W(r_{ij}, h)] \approx \sum_j f_j \nabla W(r_{ij}, h),
\end{equation}
where $\nabla W(r, h)$ can be computed analytically.
This definition fails to converge with decreasing smoothing length, and therefore has a zeroth order error. Other definitions that limit this error can be developed \citep{Price2012}.

In order to correct for this error, \citet{Lanson2008a,Lanson2008b} introduced a renormalized mesh-free derivative that is second order accurate in $h$.
Given the function $f(\myvec{x})$, its gradient is defined as
\begin{equation} \label{eq:LV_gradient}
\nabla f(\myvec{x}) = \sum_{j} \omega_j \left[ f(\myvec{x}_j) - f(\myvec{x}) \right] \tensor{B}(\myvec{x}) \myvec{\mu} (\myvec{x}, \myvec{x}_j),
\end{equation}
where the matrix $\tensor{B} = \tensor{E}^{-1}$ and
\begin{equation} \label{eq:E_matrix}
E^{\alpha \beta} (\myvec{x}) = \sum_{j} \omega_j (x^\beta_j - x^\beta) \mu^\alpha (\myvec{x}, \myvec{x}_j).
\end{equation}
The \textit{kernel vector} $\myvec{\mu}$ must be an antisymmetric, continuous and bounded function with compact support.
This is equivalent to the $\nabla W(r_{ij})$ terms of SPH in equation~(\ref{eq:SPH_grad}).
In that case, the contribution of the $j$-th particle to the gradient at the $i$-th position is aligned to the distance vector $\myvec{r}_{ij}$.
With the definition in equation~(\ref{eq:LV_gradient}), this is no longer the case.
Now each contribution is not necessarily aligned with the distance vector.
Furthermore, the gradient is computed based on relative differences instead of sums, contrary to equation~\ref{eq:SPH_grad}.
These properties, together with the definition of $\tensor{B}$, yield an improved mesh-free derivative without zeroth-order errors.

This gradient estimator was used in \citet{Gaburov2011} and \citet{Hopkins2015}, with the following modification.
In equation~(\ref{eq:LV_gradient}), instead of $\omega_j$, $\omega_i$ was used for convenience, as it can be incorporated within the kernel term as $\psi_j (\myvec{x}_i)$.
In the original formulation, this weight should be $\psi_i (\myvec{x}_j)$.
In both cases, the formulation yields second order accuracy.

We find the definition of \citet{Lanson2008a} more consistent with the definition of volume.
Comparing with SPH, using $\psi_j (\myvec{x}_i)$ would mean weighting each contribution to $\nabla W|_{x_i}$ by $m_i/\rho_i$, rather than the standard $m_j/\rho_j$ term \citep{Rosswog2009}.
Noticeable differences between the two weighting schemes would appear near discontinuities, where the smoothing length is no longer smoothly varying.
We leave a careful analysis of the weighting schemes to future work.

For the gradient estimator to work, matrix $\tensor{E}$ must be invertible.
This condition is fulfilled in the majority of cases, and only highly pathological particle distributions (e.g., all neighbours aligned in a 2D plane in a 3D simulation) could break this hypothesis.
To prevent matrix $\tensor{E}$ to become non-invertible, a check on the anisotropy of the local distribution of neighbours has been introduced, based on increasing the smoothing length, thus the number of neighbours, where needed (more details are given in section~\ref{sec:smoothing_computation}).

It is important to note that the kernel function defined for the gradient estimator can be different from that defined for the partition of unity.
\cite{Gaburov2011} and \cite{Hopkins2015} choose $\myvec{\mu}(\myvec{x}_i, \myvec{x}_j) =  (\myvec{x}_j - \myvec{x}_i)W(r_{ij})$, but as long as it is antisymmetric, the gradient estimator will work as expected \citep{Lanson2008a}.

Equation~(\ref{eq:LV_gradient}) can be substituted into equation~(\ref{eq:weak_solution_partition}), obtaining
\begin{align}
&\int_{\mathbb{R}^+} \sum_i V_i \left[ \myvec{U}_i \frac{\mathrm{d}\varphi_i}{\mathrm{d}t} + \myvec{F}_i \sum_{j} \omega_j \left[ \varphi_j - \varphi_i \right] \tensor{B}_i \myvec{\mu} (\myvec{x}_i, \myvec{x}_j) - \myvec{S}_i \varphi_i \right]\mathrm{d}t= \nonumber\\
&\int_{\mathbb{R}^+} \sum_i V_i \Bigg[ \myvec{U}_i \frac{\mathrm{d}\varphi_i}{\mathrm{d}t} - \nonumber\\
&\hspace{0.25cm} \varphi_i  \sum_j \big( \omega_j \myvec{F}_i \tensor{B}_i \myvec{\mu} (\myvec{x}_i, \myvec{x}_j) - \omega_i \myvec{F}_j \tensor{B}_j \myvec{\mu} (\myvec{x}_j, \myvec{x}_i) \big)
   - \myvec{S}_i \varphi_i \Bigg]\mathrm{d}t\,.
\end{align}
Following the general idea of the Godunov scheme \citep[see, for example,][for details]{LeVeque1992} the fluxes $\myvec{F}_i$, $\myvec{F}_j$ can be substituted by $\myvec{F}_{ij}$, which is the solution of the Riemann problem with states $\myvec{U}_i$, $\myvec{U}_j$.
The exact Riemann solver from \citet{Toro2009} is used.\footnote{The implementation is similar to that of \gizmo, also based on that of \arepo.}
The scheme reads
\begin{align}
\int_{\mathbb{R}^+} \sum_i  \left[ V_i \myvec{U}_i \frac{\mathrm{d}\varphi_i}{\mathrm{d}t} -
\varphi_i  \sum_j \myvec{F}_{ij} \myvec{A}_{ij}
- V_i \myvec{S}_i \varphi_i \right]\mathrm{d}t = 0\,, \label{eq:weak_solution_gradient}
\end{align}
where $\myvec{A}_{ij} =  V_i \omega_j \tensor{B}_i \myvec{\mu} (\myvec{x}_i, \myvec{x}_j) -  V_j \omega_i \tensor{B}_j \myvec{\mu} (\myvec{x}_j, \myvec{x}_i)$
is the \textit{face vector} along which the Riemann problem is solved.
The face is located at the mid-point between $\myvec{x}_i$ and $\myvec{x}_j$, $(\myvec{x}_i + \myvec{x}_j)/2$, and is moving with velocity $\myvec{v}_\text{frame}=(\myvec{v}_i + \myvec{v}_j)/2$.
Although the face is located in between particles, its normal vector is not necessarily parallel to the direction joining them.
The states must be boosted to the new frame of reference and projected to the face, and the resulting fluxes de-boosted from the frame of reference.\footnote{'Boosting' here means a uniform-motion (with $v_\text{frame}$) plus translation coordinate transformation; i.e., a special case of a Galilean transformation.}
This is done as \citet{Hopkins2015} and \citet{Springel2010}.

The face velocity in the direction of $\myvec{A}_{ij}$ can be arbitrarily changed to follow the deformation of the volume elements.
For instance, the face velocity can be set to follow the Lagrangian motion of the fluid, which gives a zero mass flux across the face itself.
In other words, the face is moving with the contact discontinuity of the Riemann problem, at a speed $S_*$, and the mass of the tracers is kept constant.
This defines the MFM method \citep{Hopkins2015}.
On the other hand, if the face velocity does not equal the contact discontinuity velocity, mass fluxes are allowed among particles, which gives the MFV method.

Solving the Riemann problem with the (boosted) states $\myvec{U}_i$, $\myvec{U}_j$ only provides first order accuracy in space.
In order to reach second order accuracy, the states must be extrapolated onto the face on which the Riemann problem is solved.
The extrapolation will be detailed in section~\ref{sec:limiter}.

\subsubsection{Time integration} \label{sec:time_discretization}

In order to fully specify the numerical scheme, the time integration of equation~(\ref{eq:weak_solution_gradient}) must be discretized.
The first term of equation~(\ref{eq:weak_solution_gradient}) can be integrated by parts, obtaining:
\begin{equation}
\int_{\mathbb{R}^+} \sum_i \varphi_i \left[
\frac{\mathrm{d} (V_i \myvec{U}_i)}{\mathrm{d}t} +
\sum_j \myvec{F}_{ij} \myvec{A}_{ij}
- V_i \myvec{S}_i
\right]\mathrm{d}t = 0.
\end{equation}
As neither the volume nor the test function can be zero, the equality is true if and only if
\begin{equation}
 \frac{\mathrm{d} (V_i \myvec{U}_i) }{\mathrm{d}t} +
   \sum_j \myvec{F}_{ij} \myvec{A}_{ij}
   = V_i \myvec{S}_i.  \label{eq:semi_discretised}
\end{equation}
The above equation means that the quantities $\sum_i V_i \myvec{U}_i = (M, \myvec{P}, E)$ are conserved by construction as long as the source term $\myvec{S}=0$ and $\myvec{F}_{ij}\myvec{A}_{ij}=-\myvec{F}_{ji}\myvec{A}_{ji}$.
The latter condition can be shown to be true for the face vector definition given in the previous section.

A second order operator splitting method is used for the time integration.
We discretize the temporal derivative for the conservation equations using a second order mid-point scheme such that the fluid state vector at step $n+1$, before including any contribution from sources, is
\begin{equation}
   \tilde{\myvec{U}}_i^{n+1} = \myvec{U}_i^n - \Delta t \sum_j \myvec{F}_{ij}^{n+1/2}\myvec{A}_{ij}, \label{eq:hydro_discr}
\end{equation}
where $\Delta t$ is the discrete timestep, and the fluxes are computed at step $n+1/2$ (i.e., extrapolated in time by $\Delta t/2$).
The extrapolation is done as in \citet{Hopkins2015} and \citet{Springel2010}.
Furthermore, the source terms are also extrapolated half a step, e.g., gravity terms or those arising from the comoving integration.

Source terms are added to the above state vector by averaging them over one step as follows:
\begin{equation}
\myvec{U}_i^{n+1} = \tilde{\myvec{U}}_i^{n+1}  + \frac{\Delta t}{2}(\tilde{\myvec{S}}_i^{n+1} + \myvec{S}_i^{n}), \label{eq:source_term_discr}
\end{equation}
where $\tilde{\myvec{S}}_i^{n+1}=S_i(\tilde{\myvec{U}}_i^{n+1})$.

The above discretization assumes that $\Delta t$ is global, and all particles evolve with the same timestep.
However, in simulations evolving a large range of physical scales, it is  more numerically efficient to adapt the timesteps to the physical state of individual particles.
This is widely done by allowing a power-of-two hierarchy of timesteps, where individual timesteps are defined by $\Delta t_i = 2^{-r_i} \Delta t_0$, with $\Delta t_0$ being the maximum allowed timestep, and $r_i$ a positive integer (called \textit{rung}).
The criteria to compute individual timesteps are given in section~\ref{sec:timesteps}.

In order to maintain explicit conservation when applying fluxes between particles on timesteps of different length, we followed the implementation of \citet{Springel2010}.
When adding the $\myvec{A}_{ij} \myvec{F}_{ij}$ contribution to the conserved variables, the minimum of the two timesteps, $\Delta t = \min[\Delta t_i, \Delta t_j]$, is used.
This contribution is then added to both particles, such that the interactions are always symmetric.
Further details can be found in \citet{Springel2010}, section 7.2.

\subsubsection{Smoothing length estimator} \label{sec:smoothing_computation}

In simulations with high density contrast it is unfeasible to use a fixed value for the smoothing length, as resolution will be kept constant although the density of tracers is not.
This justified the introduction of variable smoothing lengths in simulations, such that the resolution is enhanced where needed.

It is usual to define the smoothing length such that it encloses an approximate number of neighbours, $N_\text{NGB}$,\footnote{The number of neighbours can indeed be a non-integer. Other parametrisations of the same equation make it explicit by introducing the parameter $\eta$, which defines the smoothing length as function of the mean interparticle spacing  \protect\citep{Dehnen2012,Price2012}.} assuming constant density inside the compact support,
\begin{equation}
\frac{4\pi}{3} n_i h_i^3 = N_\text{NGB},
\end{equation}
where $n_i$ is an estimate of the local number density, $\omega_i^{-1}$ in this work (other SPH implementations may use $\rho_i/m_i$).
This equation, together with $n_i$ itself, form an implicit set of equations that is solved iteratively.

The requirement mentioned in section~\ref{sec:gradient_estimator}, is that the tensor $\tensor{E}$ is invertible.
From a mathematical point of view, it is unless particles are distributed regularly on a plane or along a line, in a three-dimensional domain.
From a numerical point of view, although the tensor can be inverted, any particle configuration close to perfect ordering in two or one dimensions can make the gradient estimate very noisy.
Expanding the example of particles on a plane in a three-dimensional domain, having just one neighbour off the plane, would make $\tensor{E}$ invertible.
However, along the direction perpendicular to the plane, the gradient would be unreliable, as only one particle is sampling the third dimension.
One way to alleviate this is to increase the neighbours search radius for including more off-plane particles, if the particle distribution within the compact support is not sufficiently isotropic.

The (an)isotropy of the particle distribution within the compact support can be measured using the condition number, $N_\text{cond}$, as in \citet{Hopkins2015}.
It is defined as follows,
\begin{equation}
   N_{\text{cond},i} = \nu^{-1} \left( ||\tensor{B}_i||\, ||\tensor{E}_i|| \right)^{1/2},
   \label{eq:Ncond}
\end{equation}
where the norm $||\tensor{A}|| = \sum^{\nu}_{\alpha,\beta=1} (A^{\alpha \beta})^{2}$.
The condition number is unity when $\tensor{E}_i \propto \tensor{I}$, meaning a perfectly symmetric particle distribution.
It tends to infinity in pathological configurations, when $\tensor{E}$ is close to be singular.
A minimum of $N_\text{cond}=100$ is set as threshold to capture pathological distributions of particles.
If a particle fails the check, $N_\text{NGB}$ is increased by a factor $1.2$, and the iterative process repeated.
We find that this is sufficient to avoid too noisy gradient estimates that could invalidate the solution.

We note that when the smoothing length changes significantly within a few particles, e.g., near contact discontinuities, the accuracy of the estimate of the volume partition drops to first order in space.
This is not of major concern, as the solver itself is of first order near discontinuities (see section~\ref{sec:limiter}).

\subsubsection{Spatial extrapolation and limiters} \label{sec:limiter}

If the cell-centered values of the fluid variables were used to solve the Riemann problem, the resulting scheme would be only first order accurate in space \citep{LeVeque1992}.
One way to obtain a second order scheme is to extrapolate in space the states to the face where the Riemann problem is to be solved.
This requires knowledge of the primitive variables' gradients.

Equation~(\ref{eq:LV_gradient}) is a second order accurate gradient estimator that can be used to extrapolate the states to the face location as
\begin{equation}
\myvec{U}_L = \myvec{U}_i + \frac{1}{2}(\myvec{x}_j - \myvec{x}_i)\nabla \myvec{U}_i\,,
\end{equation}
and conversely, in the other direction,
\begin{equation}
\myvec{U}_R = \myvec{U}_j + \frac{1}{2}(\myvec{x}_i - \myvec{x}_j)\nabla \myvec{U}_j\,.
\end{equation}

This yields a second order accurate scheme that, however, can lead to oscillatory behaviours near discontinuities \citep{Barth89}.
A common solution to damp these oscillations is to locally reduce the order of the scheme.
This can be done simply by using a gradient limiter as
\begin{equation}
\nabla U^\beta_i = \alpha^\beta_i \nabla U^\beta_i\,,
\end{equation}
where $\alpha^\beta_i \in [0,1]$ will determine whether we apply a second order scheme for the $\beta$ variable ($\alpha=1$) or a first order scheme ($\alpha=0$).
The key problem is to determine this limiting factor such that it is $\alpha=0$ close to discontinuities and $\alpha=1$ in smooth flows.

\citet{Barth89} devised a limiter where $\alpha$ is defined such that the extrapolated values must be bounded by the maximum/minimum values of the neighbours.
This avoids the introduction of spurious maxima/minima that lead to oscillatory behaviours.
However, their criterion can overlimit gradients even in smooth flows.
Therefore, we opted for the \textit{conditioned} \citet{Barth89} limiter, following \citet{Hopkins2015} (their equation B3),
where the maximum (minimum) allowed extrapolated values are increased (decreased) if the particle distribution is close to be isotropic.
The isotropy is established with equation~(\ref{eq:Ncond}).
We also implemented a general pair-wise limiter \citep[equation B4 of][with $\psi_1=0.5$, $\psi_2=0.25$]{Hopkins2015}.
Without the latter, most of the test simulations including gravity source terms failed to give the correct solution.

\subsubsection{Internal energy/entropy integration} \label{sec:triple_energy}

The hydrodynamics equations (equation~\ref{eq:semi_discretised})
evolve mass, momentum and total energy of the particles, $\myvec{U}_i=(m_i, \myvec{P}_i, E_i)$.
These are globally conserved as long as the source terms, $\myvec{S}_i$, are null.
At the end of each timestep, the primitive variables, $\rho_i$, $\myvec{v}_i$ and $p_i$ are computed from $\myvec{U}_i$, which has been updated by the hydrodynamic scheme (equation~\ref{eq:hydro_discr}) and the source terms (equation~\ref{eq:source_term_discr}).

In the specific case of pressure,
\begin{equation}
   p_i = \left(E_i - \frac{\myvec{P}_i\cdot\myvec{P}_i}{2m_i}\right) \frac{\gamma -1}{\omega_i},
   \label{eq:pressure_tot}
\end{equation}
where $\omega_i$ has been computed at the end of the timestep.
In the subtraction of the kinetic energy, when $E_i \simeq E_{\mathrm{kin},i}$, the calculation of the pressure can be contaminated by numerical noise \citep{Bryan1995}.
Although in this regime pressure forces are not dominant in the dynamics, a noisy estimate of pressure (and consequently of internal energy) can have a significant effect when non-hydrodynamic processes are taken into account, e.g., radiative cooling.

To prevent this, the internal energy is integrated in parallel to the total energy \citep{Gaburov2011},
\begin{equation}
   \frac{\mathrm{d}U_i}{\mathrm{d}t} = \frac{\mathrm{d}E_i}{\mathrm{d}t} - \myvec{v} \cdot \frac{\mathrm{d}\myvec{P}_i}{\mathrm{d}t} + \frac{\myvec{v}^2}{2} \frac{\mathrm{d} M}{\mathrm{d}t},
   \label{eq:internal_energy_integration}
\end{equation}
where the derivatives are those computed from the hydrodynamics equations.
Then, the pressure is simply $p_i=(\gamma -1)U_i\omega_i^{-1}$.
This way, total energy is no longer conserved due to integration errors, but a smoother pressure field is recovered in kinetically dominated flows.

Although integrating the internal energy gives a more precise evolution of the pressure, there are still regimes where numerical noise can develop,
namely, when $\mathrm{d}E_i/\mathrm{d}t \approx \myvec{v} \cdot \mathrm{d}\myvec{P}_i/\mathrm{d}t$.
We have found this in regimes where the fluid is cold and falling into a gravitational potential, such as in the Zel'dovich pancake (section~\ref{sec:zeldovich}).
Under this condition, particles can get artificially heated during the collapse, with their entropy increasing way before a shock is formed.
However, entropy-conserving SPH formulations are not affected by this by construction.
There, artificial viscosity must be added to increase the particles entropy in shocks \citep{Rosswog2009}.
In addition, two-body encounters between collisional and collisionless particles (e.g., gas and dark matter particles in cosmological simulations) can artificially heat the gas \citep{Steinmetz1997}.

One approach to mitigate these problems in mesh-based methods is to, under certain conditions, assume that the fluid is perfectly entropic, such that it can be evolved adiabatically \citep{Ryu1993}.
We improved the evolution of pressure in gravitationally dominated, cold, smooth flows by following the method in section~3.5 of \citet{Springel2010}.
Whilst we follow their method for the MFV scheme, for the MFM we took advantage of the Lagrangian formulation of the scheme, where $\mathrm{d}S_i/\mathrm{d}t=0$, to simplify the algorithm.
When entropy is evolved, the pressure is given by $p_i = S_i \omega_i \rho_i^{\gamma -1}$.
In summary, we joined the dual-energy evolution \citep{Bryan1995} and the entropy switch \citep{Ryu1993}.
All are evolved in parallel.
At the end of each timestep, the most appropriate is used for updating the pressure, according to the following criteria.

To check whether the flow is smooth, the maximum relative kinetic energy of neighbours,
$E_{\mathrm{kin},i}^\mathrm{max}$,
is gathered.\footnote{This can be done when, for example, computing the smoothing length, such that it does not require any additional neighbours search.}
Kinetic and gravitational energies are computed as $E_{\mathrm{kin},i} = \myvec{P}_i\cdot\myvec{P}_i/(2m_i)$ and $E_{\mathrm{grav},i}= m_i|\myvec{a}_\mathrm{grav}|h_i$, respectively.
With this information:
\begin{enumerate}
   \item if $U_i>\alpha_1 (E_{\mathrm{grav},i} + E_{\mathrm{kin},i})$, the pressure can be recovered from the total energy; else
   \item if $U_i<\alpha_2(E_{\mathrm{kin},i}^\mathrm{max} + U_i)$ or $U_i < \alpha_3 E_{\mathrm{grav},i}$, the flow is cold and smooth and it is assumed adiabatic, thus $p_i$ is computed from the entropy;
   \item otherwise, the pressure is computed from the internal energy.
\end{enumerate}

We took as default values $\alpha_1 = 1/100$ and $\alpha_2, \alpha_3=1/1000$.
It is possible to use other criteria based on sound speed instead of energy, as in \citet{Springel2010}.
We have found our choice simple enough to deliver satisfactory results in the three regimes, without increasing the computational time significantly.

In \cite{Hopkins2015}, the internal energy is evolved by default, and an entropy switch is activated when a cool, smooth flow is detected.
However, there is no switch to compute the pressure from the total energy, that we find necessary to recover manifest conservation in our hydrodynamic tests.

We must note one caveat: the above criteria (specifically the first) is not Galilean invariant.
This is due to the use of $E_{\mathrm{kin},i}$, which changes under a velocity boost of the frame of reference.
This is expected because, even though the scheme is invariant, the numerical errors that contaminate the pressure field are not.

When performing cosmological simulations with a UV background that maintains the internal energy above some minimum value, we found that the entropy switch is rarely activated.
For these simulations, to save computing time and memory, we disable the entropy switch, such that only the internal and total energy are integrated.

\subsubsection{Gravity} \label{sec:Gravity}

One of the advantages of \pkdgravthree is that the gravity is computed very efficiently, using GPU acceleration whenever possible.
We have avoided major modifications to the gravity module, and we just reused some of the information provided by the solver, such as the gravitational acceleration.
We give only a brief summary of this part of the code.

\pkdgravthree solves the Poisson equation using the Fast Multipole Method \citep[FMM,][]{Greengard1997}, effectively scaling as $\mathcal{O} (N)$ \citep{Potter2017}.
The solver is highly optimized for vectorized instructions, such as SSE and AVX, and uses the GPU for particle-particle interactions, multipole evaluations and periodic boundary conditions \citep[using the Ewald's summation method, ][]{Hernquist1991}.

To include the gravitational forces in our implementation of the hydrodynamic solver, we follow \citet{Springel2010}, section 5.4.
The gravitational work exerted on the particle in one timestep $\Delta t$ is
\begin{equation}
   \Delta E_{\mathrm{grav},i} = m_i \myvec{v}_i \cdot \myvec{a}_i \Delta t_i + \frac{1}{2} \myvec{a}_i \cdot \sum_j \Delta m_{ij} \myvec{r}_{ij}\,,
   \label{eq:gravity_source}
\end{equation}
where $\Delta m_{ij}$ is the mass exchanged between the $i$-th and $j$-th particles in the hydrodynamic solver, which is zero in the case of the MFM scheme.
The change in momentum is simply $\Delta \myvec{P}_{\mathrm{grav},i} = m_i \myvec{a}_i \Delta t_i$.
In both cases, the source terms are integrated using equation~(\ref{eq:source_term_discr}).

We note that we have not (yet) implemented adaptive softening.
Instead, we set a global softening length, as usually done for simulations of galaxy formation in cosmological volumes.
It is expressed in comoving coordinates, and a maximum can be set in physical coordinates.
Furthermore, constant per-particle softening lengths can be set (e.g., for zoom-in simulations).

\subsubsection{Comoving integration} \label{sec:comoving}

The case in which the fluid equations are solved in an expanding universe is of special interest, as the main application of this code are simulations of galaxy formation.
In an evolving expanding universe with a time-dependent scale factor, $a(t)$, the fluid equations are solved in the comoving frame of reference.
Physical coordinates, $\myvec{x}(t)$, are related to comoving coordinates, $\hat{\myvec{x}}(t)$, through the relation $\myvec{x}(t) = a(t) \hat{\myvec{x}}(t)$.
The physical velocity is then $\myvec{v}(t)\equiv\dot{\myvec{x}}(t) = \dot{a}(t) \hat{\myvec{x}}(t) + \hat{\myvec{v}}(t)$, where $\hat{\myvec{v}}(t) = a(t) \dot{\hat{\myvec{x}}}(t)$ is the peculiar velocity, whereas the other term accounts for the Hubble expansion.
Throughout the rest of the derivation, we will omit the time dependence for clarity.

Generally, the comoving density is written as $\hat{\rho} = \rho a^3$, as this definition recover the continuity equation without source term.
The definition of the other thermodynamic variables can be somehow arbitrary, as long as the hydrodynamics equations are derived and solved consistently.
This has lead to various conventions.
For entropy conserving SPH codes, the comoving entropy is set equal to the physical, $\hat{s} = s$, which leads to a pressure and internal energy dependence on the scale factor as $\hat{p} = p a^{3\gamma}$ and $\hat{u}=ua^{3(\gamma-1)}$, when combined with the above definition of the comoving density.

In our implementation, entropy is rarely used, and we do not see any advantage in using this convention.
Instead, we set $\hat{u}=u$, which leads to $\hat{p} = p a^3$, as in general either the total or internal energies are used.
Under this definitions, the fluid equations (including gravity) become:
\begin{align}
   &\frac{\partial \hat{\rho}}{\partial t} + \frac{1}{a} \hat{\nabla} (\hat{\rho} \hat{\myvec{v}}) = 0\,, \\
   &\frac{\partial \hat{\rho} \hat{\myvec{v}} }{\partial t} + \frac{1}{a} \hat{\nabla} (\hat{\rho} \hat{\myvec{v}}\otimes\hat{\myvec{v}} + \hat{p} \tensor{I}) = -  \frac{\dot{a}}{a} \hat{\rho} \hat{\myvec{v}}  - \frac{\hat{\rho}}{a} \hat{\nabla} \hat{\phi}\,, \\
   &\frac{\partial \hat{\rho} \hat{e}}{\partial t} + \frac{1}{a} \hat{\nabla} \left[(\hat{\rho} \hat{e} + \hat{p})\hat{\myvec{v}} \right] = -\frac{\dot{a}}{a} (\hat{\rho} \hat{v}^2 + 3 \hat{p}) - \frac{\hat{\rho} \hat{\myvec{v}}}{a} \hat{\nabla} \hat{\phi}\,,
\end{align}
where the peculiar gravitational potential is $\hat{\phi} = \phi + \frac{1}{2} a \ddot{a} \hat{x}^2$ and $\hat{\nabla}$ denotes the gradient with respect to comoving coordinates.
Additionally, the evolution of the internal energy takes the form
\begin{equation}
   \frac{\partial \hat{u}}{\partial t} + \frac{1}{a}(\hat{\myvec{v}}\cdot \hat{\nabla})\hat{u} + \frac{1}{a} \frac{\hat{p}}{\hat{\rho}} \hat{\nabla} \hat{\myvec{v}} = - 3 \frac{\dot{a}}{a} \frac{\hat{p}}{\hat{\rho}}\,.
\end{equation}

In the comoving frame of reference, two major changes to the fluid equations can be seen:
(1) all spatial derivatives are scaled by $1/a$, which is taken into account as an increase of the fluxes computed by the Riemann solver, and
(2) source terms proportional to the Hubble parameter, $H=\dot{a}/a$, appear, and are treated with a modified version of equation~(\ref{eq:source_term_discr}):
\footnote{This modification can be applied only because the source terms are functions of the fluid variables.
For the energy conservation equation, this only happens if $\gamma=5/3$.}
\begin{equation}
U^{n+1} = \frac{\tilde{U}^{n+1} - \eta  \Delta t H^n U^n}{1 + \eta \Delta t H^{n+1}},
\end{equation}
which is applied to the momentum ($\eta=1/2$), total energy ($\eta = 1$), internal energy ($\eta = 3/2$) and entropy ($\eta = 3/2$) equations.
The rest of the hydrodynamic scheme is left unchanged.

\subsubsection{Timestep computation} \label{sec:timesteps}

The adaptive time evolution in \pkdgravthree is performed over a hierarchical structure of timesteps.
Derived from the largest allowed timestep, $\Delta t_0$, the range of possible timesteps are power of two fractions of it: $\Delta t = 2^{-r} \Delta t_0$, where $r$ is the timestep rung.

When performing a simulation, the main loop starts with the base timestep ($r=0$).
Then, if there are particles with $r_i>r$, the step is divided in two sub-steps with a higher rung, $r \leftarrow r+1$.
This is repeated recursively until the maximum rung of the particles, $r_\mathrm{max}$, is reached.
At that point a full step with $\Delta t = 2^{-r_\mathrm{max}} \Delta t_0$ is performed.

In order to determine the timestep of particles, $\Delta t_i$, or their rung, $r_i$, we have employed three different criteria.
The first is the local Courant-Friedrichs-Levy (CFL) condition, preventing the hydrodynamic interaction from propagating farther than one compact support radius in one timestep,
\begin{equation}
    \Delta t_{\mathrm{CFL},i} = C_\mathrm{CFL} \frac{h_i}{|v_{\mathrm{sig},i}|},
\end{equation}
where
\begin{equation}
    v_{\text{sig},i} = \max_j \left[ c_{\text{s},i} + c_{\text{s},j} - \min \left( 0, \frac{(\myvec{x}_i - \myvec{x}_j)\cdot(\myvec{v}_i - \myvec{v}_j)}{|\myvec{x}_i - \myvec{x}_j|} \right) \right]
\end{equation}
is the signal velocity, and $c_{\text{s},i}$ and $c_{\text{s},j}$ are the sound speeds of the $i$-th and $j$-th particles, respectively.

The second is an acceleration criterion preventing strong changes in the velocity of particles,
\begin{equation}
\Delta t_{\text{acc},i} = C_\text{acc} \sqrt{ \frac{h_i}{a_i} },
\end{equation}
where $\myvec{a}_i = \myvec{a}_{\text{grav},i}+\myvec{a}_{\text{hydro},i}$ is the particle acceleration, and $C_\text{acc}$ is a factor controlling the accuracy.
We select the timestep rung, $r_i$, such that it is the minimum integer that fulfills
\begin{equation}
\min [\Delta t_{\text{CFL},i}, \Delta t_{\text{acc},i}] > 2^{-r_i} \Delta t_0\,,
\end{equation}
before applying the third criterion.

The third criterion limits the timestep rung of neighbouring particles with respect to the particle's rung.
This is the limiter introduced by \citet{Saitoh09}, who studied energy conservation in the presence of large particle to particle variations of internal energy.
We implemented it by forcing interacting particles to have, at most, a factor of four difference in their timesteps, $|r_i-r_j| \le 2$.

\subsubsection{Units and output format}

The code units in \pkdgravthree are defined such that the value of the gravitational constant, $G$, equals unity.
By providing mass and length units, the unit system is then fully specified.
For cosmological setups, we fix the length unit to have a computational box of size unity, and the mass unit is set such that $\rho=\Omega_m$ in code units, where $\Omega_m$ is the universe matter density parameter.
The required factors to convert from code units to physical units are appended to all snapshots, and a Python script is also provided to compute all the conversion factors.
We note that the mass and length units are not scaled by the Hubble parameter, $h = H_0/(100\kmsMpc)$, where $H_0$ is the Hubble constant.

Contrary to other codes, time integration in cosmological setups is performed over physical time, rather than scale factor.
As a result, simulation snapshots are equally spaced in time by default.
However, the code can be easily adapted to, given a list of arbitrary times (or redshifts), produce outputs at those exact instants.

\pkdgrav uses by default the TIPSY format for storing snapshots.\footnote{\url{http://faculty.washington.edu/trq/hpcc/tools/tipsy/tipsy.html}}
The format is convenient for displaying and analyzing results of N-body simulations, but however lacks the flexibility required for simulations of galaxy formation.
In these, the structure of the output can change substantially depending on the particle types and the physics being simulated.
For being able to adapt to different output fields with ease, we have chosen to update the legacy HDF5\footnote{\citet{hdf5}: \url{https://www.hdfgroup.org/solutions/hdf5/}} module of \pkdgravthree to fully support both N-body and hydrodynamic simulations.
HDF5 provides a cross-platform, consistent and straightforward way of sharing data.
Because of this, it is already extensively used in the Astrophysics community, and there are some `standard' naming conventions for the outputs of cosmological simulations, mainly based in \gadgettwo.
\pkdgravthree mostly follows this naming convention, such that its output can be easily analysed by existing tools and scripts.

%%%%%%%%%%

\section{Results}\label{sec:Results}

In order to regularly validate the code during development, we designed an automated suite of hydrodynamic test simulations.
The suite comprises a number of pure hydrodynamic simulations to test the solver without external sources, and a couple of simulations with self-gravity and external gravity.
All tests, including initial conditions and runtime parameter files, are publicly available.
We describe and discuss pure hydrodynamic tests in section~\ref{sec:hydro_tests}, and the coupling to gravity in section~\ref{sec:grav_tests}.
Lastly, in section~\ref{sec:scaling}, the scalability of the code is studied in a cosmological simulation.

\subsection{Hydrodynamic tests} \label{sec:hydro_tests}

In this section, we show the results of hydrodynamic idealised tests that are extensively used to check the correctness of the scheme and compare different codes \citep[e.g.,][]{Springel2010,  Bryan2014, Hopkins2015,  Wadsley2017, Hubber2018, Springel2021, Borrow2022}.
We note that all the tests presented hereafter are performed in three-dimensions, as we consider this the most representative configuration to be used in production runs.
In the case that the problem is formulated in 1D, we add a few ($\gtrsim 4$) planes of particles along the other dimensions, but we do not force particles to move along the direction specified by the problem nor we assume zero fluxes along the perpendicular directions.

\subsubsection{Soundwaves} \label{sec:soundwaves}

Probably the simplest non-trivial solution of the fluid equations is the propagation of soundwaves in a homogeneous background.
It has a simple analytical solution and, as it does not involve non-linear behaviours, can be used to evaluate the convergence of hydrodynamic methods.

\begin{figure}
   \centering
   \includegraphics[width=0.95\columnwidth,trim=0mm 0mm -20mm 0mm]{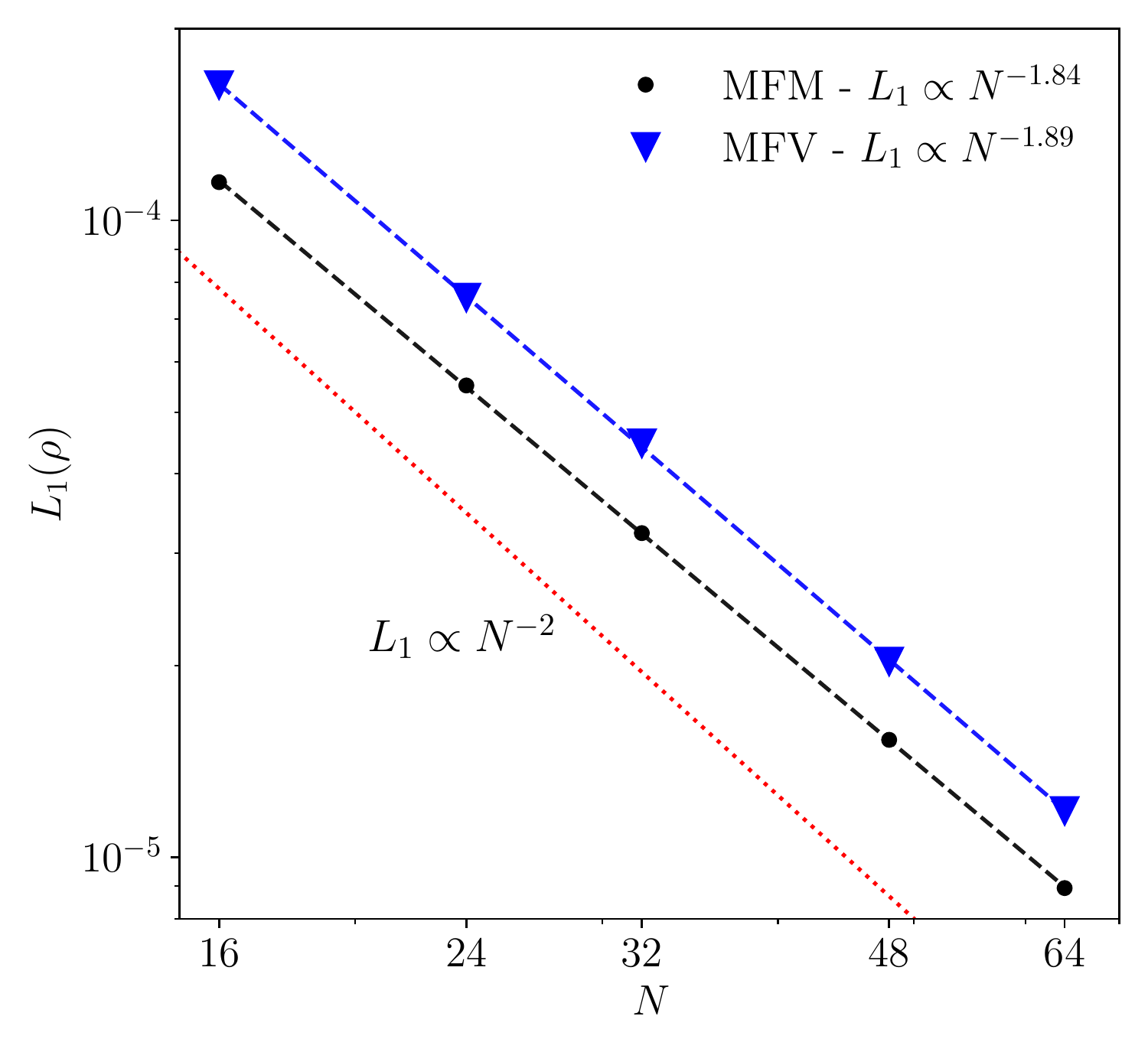}
   \caption{
      L1 norm convergence of the hydrodynamic schemes for the 3D soundwave test simulation (section~\ref{sec:soundwaves}).
      In blue triangles, the MFM scheme. In black dots, the MFV scheme.
      The dashed lines are fits to the data points, and the convergence exponent is shown in the legend.
      Both schemes are second order accurate.
   \label{fig:soundwaves}}
\end{figure}

For this test, we place a sinusoidal perturbation of amplitude $A=10^{-3}$ in a medium with unit length, density $\rho_0=1$, and speed of sound $c_s = \sqrt{\gamma}$ with $\gamma=1.4$.
The solution is then, at any time $t$,
\begin{align}
 &\rho(t) = \rho_0 ( 1 + A \cos \varphi(t) ), \\
 &p(t) = p_0 ( 1 + A \gamma \cos \varphi(t) ), \\
 &v_x(t) = A c_s \cos \varphi(t)
\end{align}
where $\varphi(t) = 2 \pi x - \omega t$ is the phase of the wave, assuming only one wavelength fits in the domain.
We increase the spatial resolution of the initial conditions in steps, by changing the number of particles along the x-axis, $N$, and consequently the thickness of the domain.
For each simulation, the $L_1$ norm of the density field is computed after one crossing time.
The convergence of the scheme is shown in figure~\ref{fig:soundwaves}.
Near to second order convergence is recovered as expected.
Furthermore, in concordance with \citet{Hopkins2015}, we find that the method is second order accurate independently of the number of neighbours.
For the figure shown above, $N_\mathrm{NGB}=32$ was used.
The gradient limiter was disabled (section~\ref{sec:limiter}), because it can degrade the convergence, in particular at low resolution.
However, we checked that the $L_1$ norms, with or without limiter, are almost indistinguishable for $N \geq 64$.

Compared with convergence tests performed with other codes, perturbation amplitudes of the order of $A \sim 10^{-6}$ or lower cannot be simulated at second order, as single precision is used for storing the mass of the particles.
This also limits the minimum $L_1$ norm to $\sim A^2$, before starting to observe physically correct non-linear behaviours \citep{Stone2008}.

\subsubsection{Sod shock tube} \label{sec:riemann}

The code ability to handle discontinuities is tested in this section.
The simplest configuration for this is a Sod shock tube, a particular case of the Riemann problem that simulates two different fluid states separated by a diaphragm that is removed at time $t=0$.
The initial values of the fluid variables, on the left and right side of the discontinuity, are $\rho_L, v_L, p_L = (1, 0, 1)$ and $\rho_R, v_R, p_R = (0.25, 0, 0.1795)$, and $\gamma = 1.4$.
These same parameters were also used in \cite{Springel2010}, \cite{Hopkins2015} and \cite{Wadsley2017}, among others.

\begin{figure}
   \includegraphics[width=0.95\columnwidth,trim=7.5mm 0mm -3.75mm 0mm]{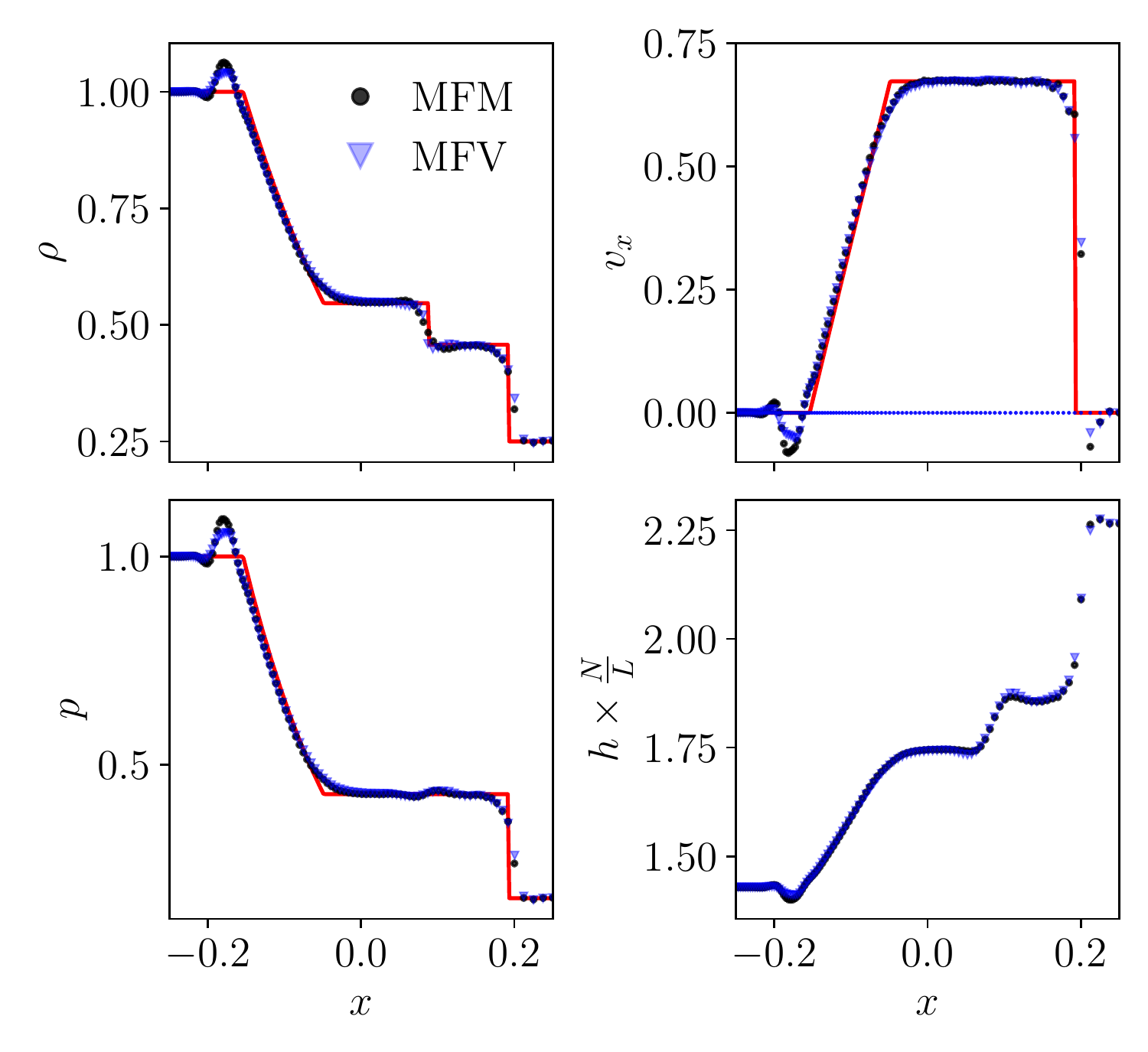}
   \caption{
      Solution of the 3D Riemann problem at $t=0.13$ for the MFM and MFV schemes, with the same color scheme as in figure~\ref{fig:soundwaves}.
      The analytical solution is shown as a red line.
      In the bottom-right panel, the smoothing length is shown, illustrating the decrease in effective resolution in the rightmost shock.
   \label{fig:riemann}}
\end{figure}

For particle-based codes, the initial density can be set either choosing the particle masses (given their fixed number density) or the particle number density (given their fixed mass).
In our case, we have chosen the latter as it better resembles the configuration that may be encountered in cosmological simulations, where density is mostly defined by particle number density.
A total of $200$ particles along the x-axis were placed, and the number of neighbours was set to $N_\textrm{NGB}=64$.
The solution at $t=0.13$ for both schemes is shown in figure~\ref{fig:riemann}.
The MFM solution is shown as black dots, whereas the MFV solution is depicted as blue triangles.
The analytical solution is shown as red lines.

Note that as the density is set by the number density, the right shock has effectively less resolution than the rarefaction wave.
This is easily seen in the bottom-right panel, where the smoothing length is shown to increase from left to right.
The density blip at $x=-0.2$ is not caused by the change of particle number density through the rarefaction wave, but by the gradient limiter we are using.
More restrictive limiters suppress this oscillation, as also mentioned by \citet{Springel2010} and \citet{Hopkins2015}.

As this test is performed in 3D, particle misalignment and perpendicular motions are plausible.
To check that this is not affecting the solution, the perpendicular velocity for the MFV simulation is shown in the top-right panel as blue dots.
This velocity is negligible and does not cause particle misalignment.
This also holds for the MFM scheme.

\subsubsection{Gresho-Chan vortex} \label{sec:gresho}

Key to the evolution of galaxies is the correct conservation of momentum in differentially rotating fluids.
To this end, the \citet{Gresho1990} vortex is a suitable test for the hydrodynamic scheme, which has also been used extensively to test other codes.
The triangular vortex is defined by the tangential velocity profile
\begin{equation}
   v_\varphi = \left\{\begin{array}{ll}
   5r &\mathrm{for}\quad r \leq 0.2 \\[2pt]
   2 - 5r &\mathrm{for}\quad  0.2<r<0.4 \\[2pt]
   0 &\mathrm{for}\quad r \geq 0.4
   \end{array}
   \label{eq:gresho}\right.
\end{equation}
and the corresponding pressure profile for hydrostatic equilibrium with $p(r=0) = 5$.
The density is initially set to unity and $\gamma = 1.4$.
The vortex Mach number at the peak is $\mathcal{M}(r=0.2) \approx 0.37$.

For the initial conditions, a Cartesian grid of particles is used.
This provides an extra difficulty to the scheme, as mixing is induced in the shear flow, and noise in the density estimate is unavoidable.
This can be mitigated by arranging particles in concentric circles, but, as this configuration is highly unlikely in a cosmological simulation, we opted for a less favourable case.

\begin{figure}
   \includegraphics[width=0.9\columnwidth,trim=0mm 0mm 0mm 0mm]{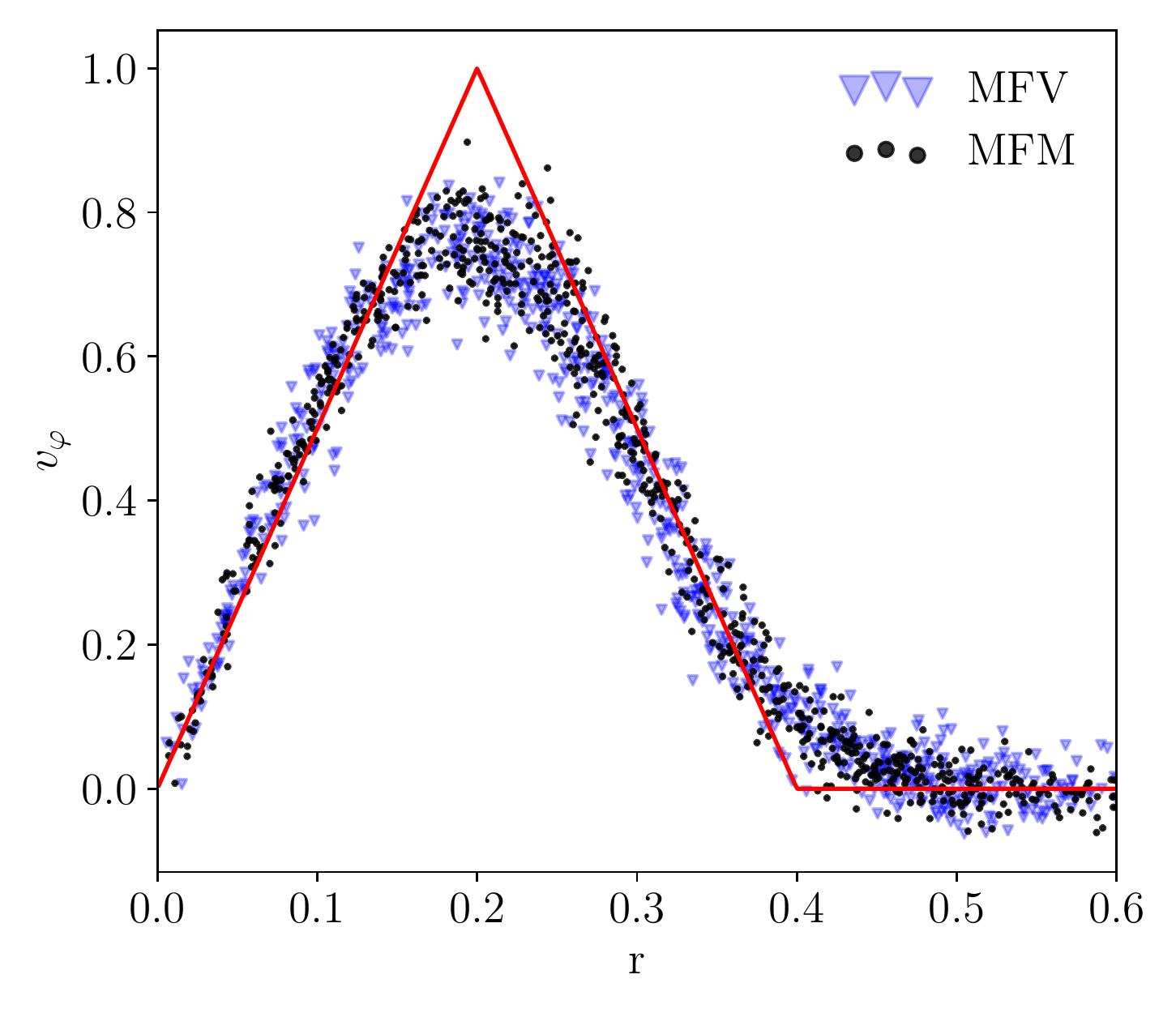}
   \caption{
      Tangential velocity for the 3D Gresho vortex (section~\ref{sec:gresho}) at $t=3$, starting from a Cartesian grid of particles.
      For clarity, only $10\%$ of the particles are plotted.
      Both solutions are in agreement and correctly recover the expected solution (continuous red line), showing only a small vortex decay.
   \label{fig:gresho_1}}
\end{figure}

The solutions for the MFM (black) and MFV (blue) schemes are shown in figure~\ref{fig:gresho_1} at $t=3$, when the peak of the vortex has performed $2.4$ orbits.
The initial velocity profile, which should be maintained in the ideal case, is shown as a red line.
In this test, $64^2$ particles were used for each plane of side $L=2$.

The extent of the decrease of the velocity peak is mostly dictated by the gradient limiter.
A more restrictive limiter produces a less noisy solution but a lower velocity peak \citep[as also seen in, e.g.,][]{Hopkins2015}.
The vortex is not destroyed and angular momentum is conserved.
Furthermore, the inner region of the vortex is still rotating as a solid body.

\subsubsection{Blob test} \label{sec:blob}

An important feature of the MFM and MFV schemes is their ability to capture contact discontinuities and instabilities without requiring any modification or tuning.
In order to test this, we performed the so-called \textit{blob} test, an hydrodynamical test that has been used in different studies to analyze the difference between hydrodynamic schemes \citep{Agertz2007,Braspenning2022}.

The blob test consist of a spherical cloud of dense, uniform gas in pressure equilibrium with a medium of lower density acting as a uniform wind.
The setup can be parametrised by the density and temperature contrast between the cloud and the wind, $\chi = \rho_\text{cl}/\rho_\text{wind} = T_\text{wind}/T_\text{cl}$, the size of the cloud, $R_\text{cl}$, and the mach number of the wind, $\mathcal{M}_\text{wind} = v_\text{wind} / c_{s,\text{wind}}$.
With these, the cloud crushing time can be obtained as $t_\text{cc} = \sqrt{\chi} R_\text{cl} / v_\text{wind}$.
Parameters and initial conditions are the same as in \citet{Braspenning2022}, such that our results can be easily compared with theirs:
$R_\text{cl}=0.1L$, with $L=1$ the vertical size of the volume, $\mathcal{M}_\text{wind}=1.5$, and $\chi$ is either 10 or 100.
The resolution is denoted by $N$, the number of particles along the vertical side of the elongated domain.
The number of particles per cloud radius is then $0.1N$.
\citet{Braspenning2022} also provide simulations with the MFM and MFV schemes described by \citet{Hopkins2015} and implemented in \swift.
We recall that our implementations differs from that of \citet{Hopkins2015}, therefore the results may not match identically.
We also point out that the development of instabilities is triggered by numerical noise rather than initial perturbations of the velocity orthogonal to the contact discontinuity.
A detailed discussion on how to trigger instabilities from the initial conditions rather than numerical noise can be found in \citet{Read2010}.

\begin{figure*}
   \centering
   \includegraphics[width=0.9\textwidth,trim=0mm 0mm 0mm 0mm]{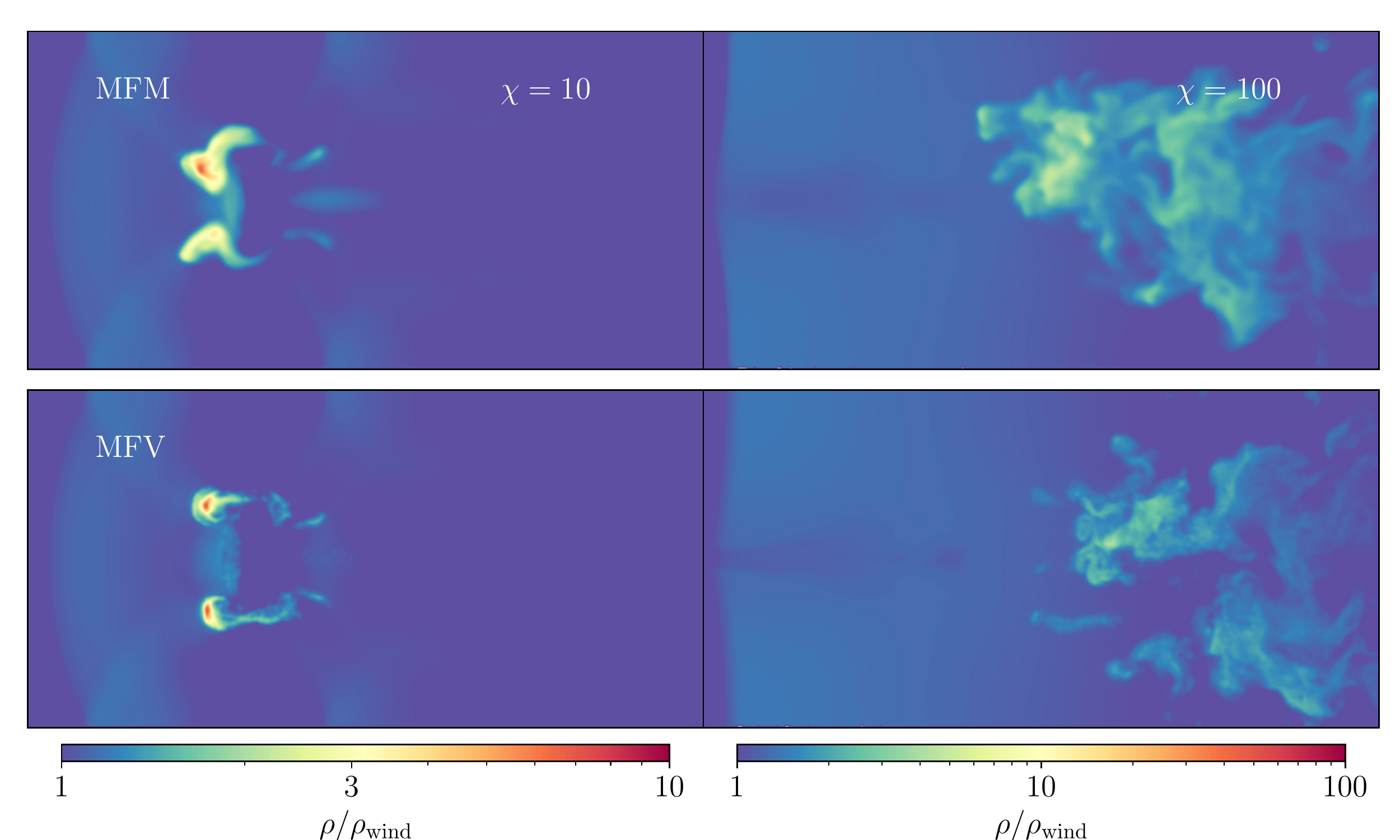}
   \caption{
       Density slices at $t=5 t_\text{cc}$ for the MFM (top row) and MFV (bottom row) schemes blob test (section~\ref{sec:blob}).
       The initial density contrast is $\chi=10$ and $\chi=100$ for the left and right columns, respectively.
   \label{fig:blob_1}}
\end{figure*}

The gas density in a slice centred at $z=0$ is plotted in figure~\ref{fig:blob_1} after five crushing times for $N=128$.
The simulations with $\chi=10$ and $\chi=100$ are shown in the left and right column, respectively.
Both schemes correctly capture the bow shock and the disruption of the gaseous sphere.
The morphology of the solution is similar for both schemes, as it is mostly dictated by the noise in the initial conditions.

\begin{figure}
   \centering
   \includegraphics[width=0.85\columnwidth,trim=0mm 0mm 0mm 0mm]{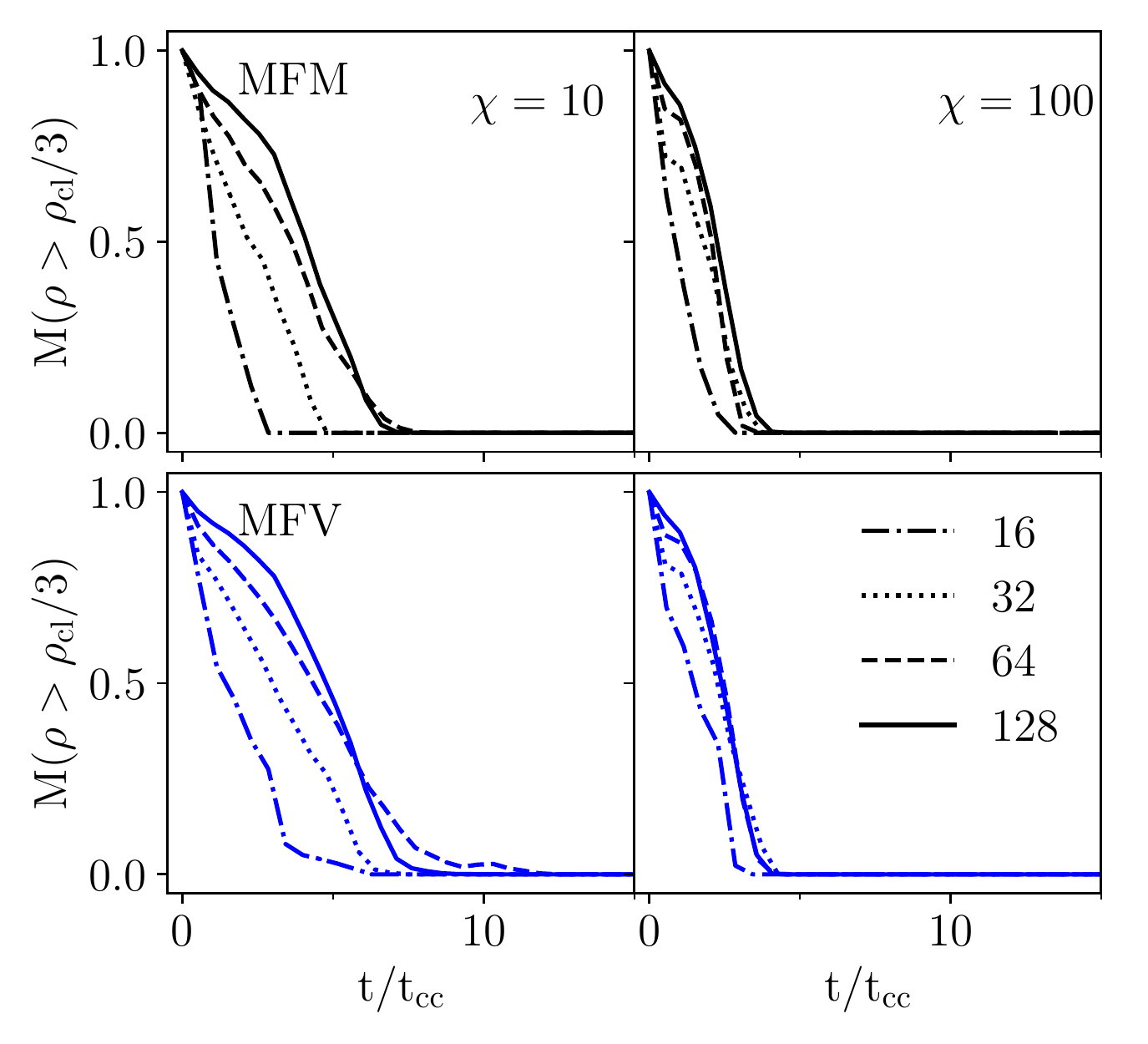}
   \caption{
       Evolution of the cloud mass at different resolutions.
       Line styles denote different values of $N$.
       The solutions using MFV and MFM schemes show similar behaviour, but the cloud can survive slightly longer in MFV.
   \label{fig:blob_2}}
\end{figure}

A quantitative measure of the disruption of the cloud if given by calculating the mass of the surviving cloud.
We do so by computing the total mass of particles with density larger than $\rho_\text{cl}/3$.
This is shown for both schemes and different resolutions ($N = [16,32,64,128]$) in figure~\ref{fig:blob_2}.

The solution for both schemes is rather similar, with the main difference being slightly longer cloud survival times for the MFV scheme at almost all resolutions.
This is compatible with the results of \citet{Braspenning2022}, but differs from those of \citet{Hopkins2015}, where both schemes show almost identical evolution of the cloud mass.
However, \citet{Hopkins2015} employs the setup and analysis of \citet{Agertz2007}, that may not be directly comparable.
In addition, the choice of slope limiter and other details of the hydrodynamic scheme can have noticeable impact on the disruption and mixing of the cloud due to instabilities.

\subsubsection{Sedov explosion} \label{sec:sedov}

The last hydrodynamic test is a point-like explosion, typically referred to as the Sedov blast-wave.
In a homogeneous medium at rest, a large amount of internal energy is deposited in a small volume, such that the internal energy per unit mass within the volume is much larger than that of the background.
A self-similar solution for the evolution of the explosion can be obtained \citep{Sedov59}.

\begin{figure}
   \includegraphics[width=0.95\columnwidth,trim=0mm 0mm -2.5mm 0mm]{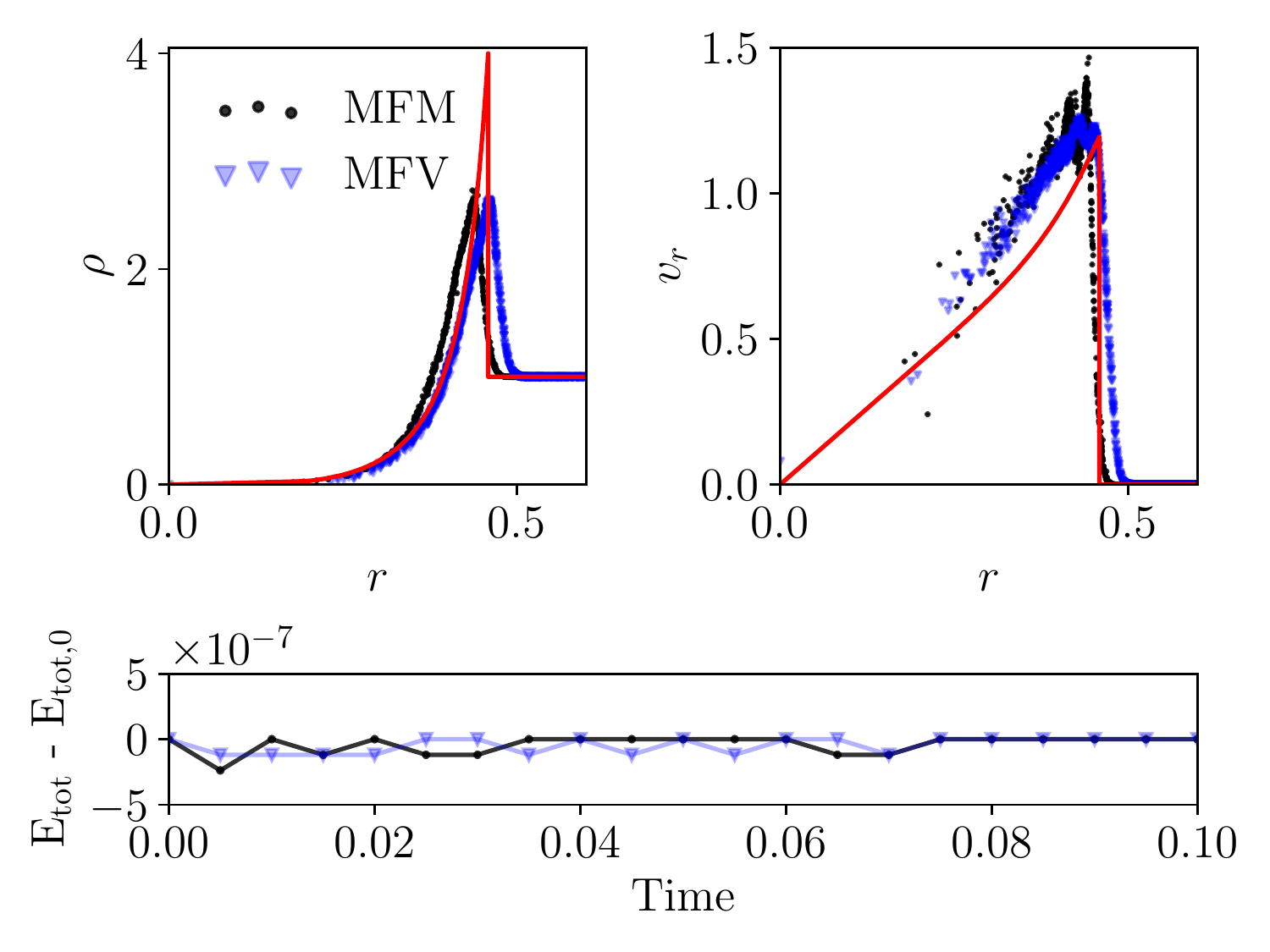}
   \caption{
      Three-dimensional Sedov blast-wave simulation at $t=0.1$ (section~\ref{sec:sedov}).
      \textit{Top panels}. Radial profiles of gas density (left) and radial velocity (right).
      The analytic solution is shown with a red line.
      Both schemes reach the same density contrast, although the shock in the MFM scheme is lagging behind the analytic solution (see text for more details).
      \textit{Bottom panel}. Total energy conservation as a function of time for both schemes.
      Energy is conserved to machine precision.
   \label{fig:sedov}}
\end{figure}

For this test, a Cartesian grid of $64^3$ particles was placed in a box of size $L=1$, with constant density $\rho = 1$.
The gas was assumed to be monoatomic, with specific heats ratio $\gamma=5/3$.
The total energy of the background medium was set to $E_0 = 10^{-5}$.
The internal energy of the central particle was then set to unity, which corresponds to an internal energy contrast of about $10^{10}$ for the chosen resolution.
The high ratio of internal energies ensures that the evolution is not affected by the background.
The density profiles at $t=0.1$ for the MFM (black) and MFV (blue) schemes are shown in figure~\ref{fig:sedov}.
The Sedov solution provides the expected density contrast at the shock front, $(\gamma + 1)/(\gamma -1)=4$.
However, with our setup, the maximum peak density is $\sim 2.7$ for both schemes.

In this test, differences between the schemes are noticeable, as it was also observed by \citet{Hopkins2015}.
Although in both simulations the total energy is conserved to machine accuracy (bottom panel of figure~\ref{fig:sedov}), the shock position is slightly behind the analytic solution position for the MFM scheme.
However, the density contrast does not seem to depend on the scheme, contrary to what was found by \citet{Hopkins2015}.
We have tried different configurations, including deactivating the slope-limiters or using a single, global timestep to evolve the system, but no configuration substantially increased the density contrast at the shock position, nor corrected the shock position for the MFM scheme.
The only viable way to increase the density contrast is increasing the resolution.
On the other hand, the delay in the propagation of the shock front seems to be inherent to the MFM scheme.
We performed several tests changing runtime parameters in order to investigate what could cause the lagging of the shock front.
We did not find any substantial improvement of the solution by arbitrarily decreasing the timestep, but we did by increasing the spatial resolution \citep[see figure 14 of][where our scheme is compared to theirs for increasing resolution]{Morton2022}.
It seems that under extreme conditions the MFM scheme falls at a lower order of accuracy than MFV,
and, as the total energy is conserved by construction, the missing kinetic energy is balanced by higher temperature within the expanding bubble.

Decreasing the energy of the explosion leads to the two solutions to agree.
This is encouraging because extreme Mach numbers ($\mathcal{M}\sim 10^5$) as in this test are not common in cosmological simulations.

\subsection{Gravity coupling tests} \label{sec:grav_tests}

The hydrodynamic solver implemented within \pkdgravthree has been coupled to the gravity solver.
Although the algorithm for computing gravity forces has not been modified, there is no unique way to couple it to the hydrodynamics, and hence we need to test our implementation.
To this end, we have performed a series of simulations including self-gravity (sections~\ref{sec:evrard},~\ref{sec:zeldovich}) and a cosmological simulation with dark matter and gas (section~\ref{sec:santabarbara}).

\subsubsection{Evrard's spherical collapse} \label{sec:evrard}

\begin{figure}
   \includegraphics[width=0.9\columnwidth,trim=-7.5mm 5mm -10mm 0mm]{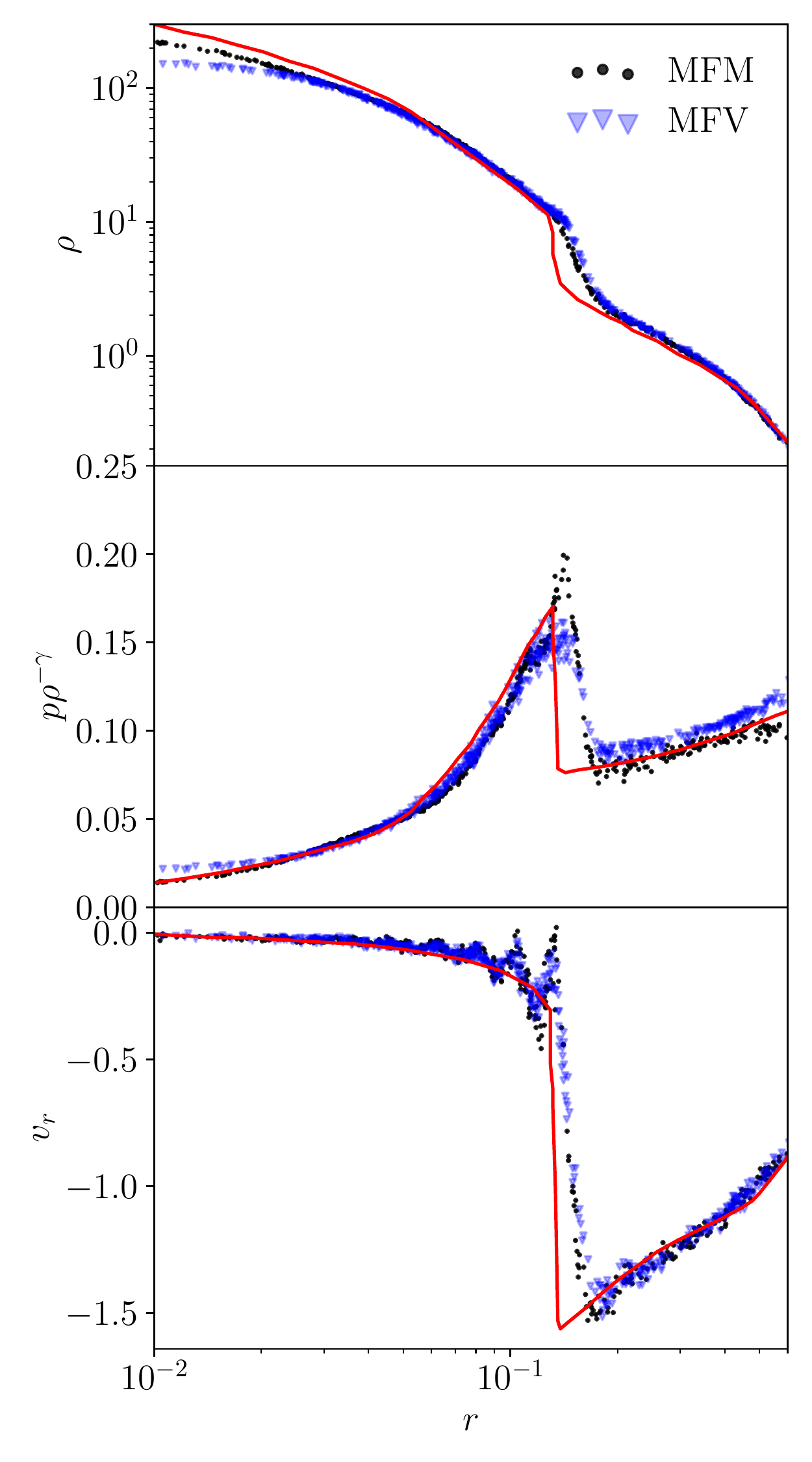}
   \caption{
      Evrard's spherical collapse of self-gravitating gas at $t=0.16$ (section~\ref{sec:evrard}).
      We show radial profiles for density (top panel), entropy (mid panel) and radial velocity (bottom panel) for the MFM and MFV schemes.
      Only $5\%$ of the particles are shown for clarity.
      The reference solution for a 1D high-order simulation is shown as red lines.
      Both schemes recover the structure of the solution although small, expected differences appear in the entropy jump at the strong shock.
   \label{fig:evrard_1}}
\end{figure}

One simulation extensively used for testing the implementation of self-gravitating gas \citep[e.g.,][]{Steinmetz1993, Springel2005b, Hopkins2015, Wadsley2017} is Evrard's spherical collapse \citep{Evrard1988}.
It consists of the collapse under self gravity of a cloud of gas with negligible initial internal energy.
During the initial phase of the collapse, potential energy is converted into kinetic.
As the inner region is compressed, its internal energy starts to increase, and a strong shock is formed.
The shock propagates outwards, stopping the infalling gas and converting kinetic energy into internal.
The end result is a gaseous sphere in hydrostatic equilibrium.
A key factor in the evolution is the correct conservation of energy, which can be carefully monitored with this test.

The initial conditions consist of a sphere with density $\rho(r) = M/(2\pi R^2 r)$ at rest, with an internal energy per unit mass of $0.05$.
The sphere has radius $R=1$, and encloses the total mass $M=1$.
The solutions at $t=0.8$ are shown in figure~\ref{fig:evrard_1} for both schemes, in a configuration with 27000 particles.
The reference solution is shown as red lines.

Both schemes correctly recover the position of the shock, and the velocity and density profiles with similar accuracy.
The main difference lies in the entropy profile (middle panel), where the entropy jump is smaller for the MFV scheme, and there is a small overestimation of the entropy ahead of the shock.

\begin{figure}
   \includegraphics[width=0.9\columnwidth,trim=0mm 5mm -5mm 0mm]{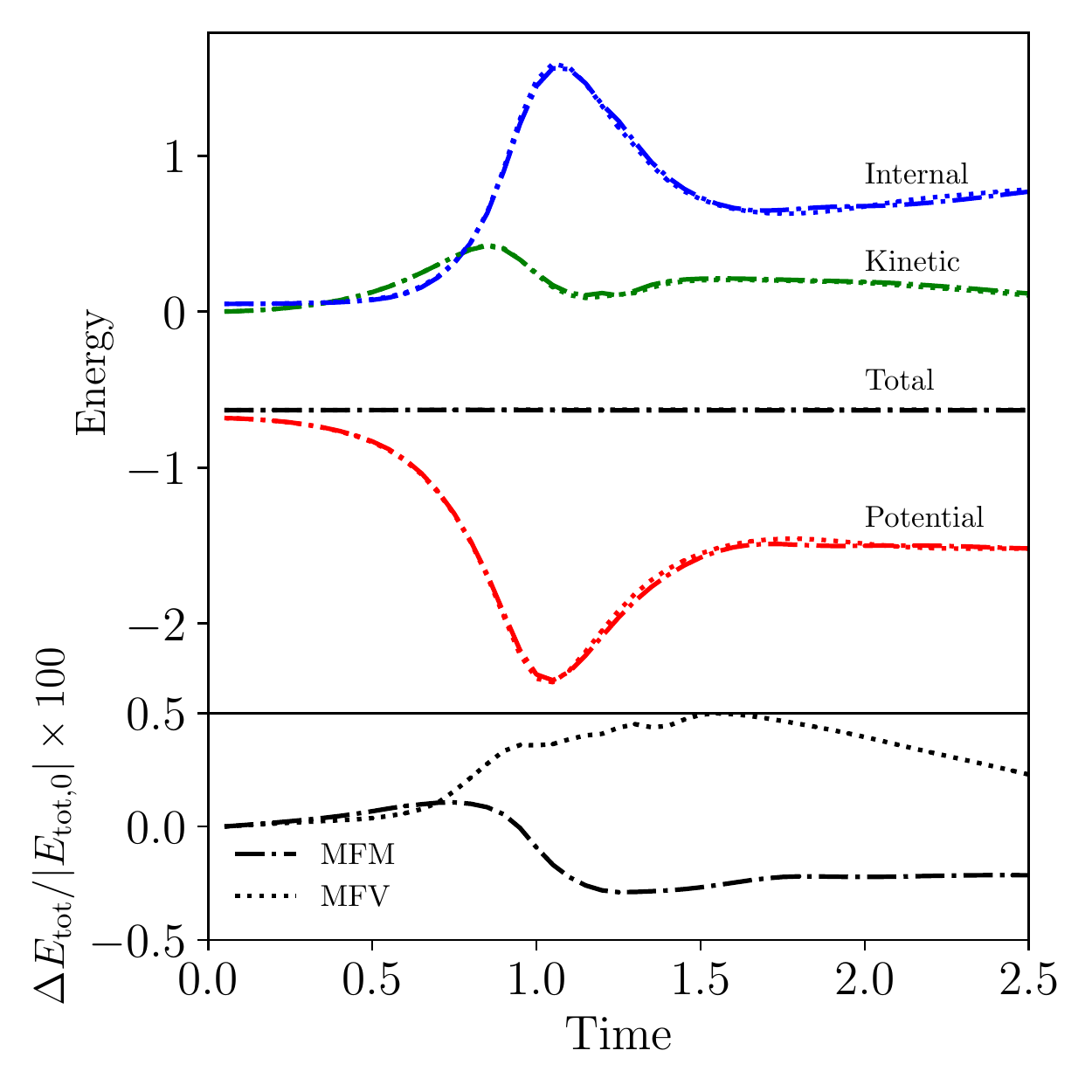}
   \caption{
      \textit{Top panel}. Time evolution of the potential (red), kinetic (green) and internal (blue) energy for the Evrard's spherical collapse (section~\ref{sec:evrard}).
      \textit{Bottom panel}. The total energy conservation is shown to be below $0.5\%$ even during the shock formation.
      The lack of perfect conservation is mainly due to errors in the calculation of gravity, rather than losses in the hydrodynamic scheme.
   \label{fig:evrard_2}}
\end{figure}

The time evolution of energy is shown in figure~\ref{fig:evrard_2}.
In the upper panel, the internal, kinetic, potential and total energy are shown.
Both schemes produce the same overall evolution of the energies.
The conservation of the total energy is presented in the lower panel.
Note that the y-axis is multiplied by a factor of $100$, and energy is conserved within $0.5$ percent.
The error in the conservation of energy is due to the gravity computation itself, rather than the hydrodynamic schemes, which by construction conserve energy to machine precision.

\begin{figure}
   \centering
   \includegraphics[width=0.9\columnwidth,trim=0mm 5mm -10mm 0mm]{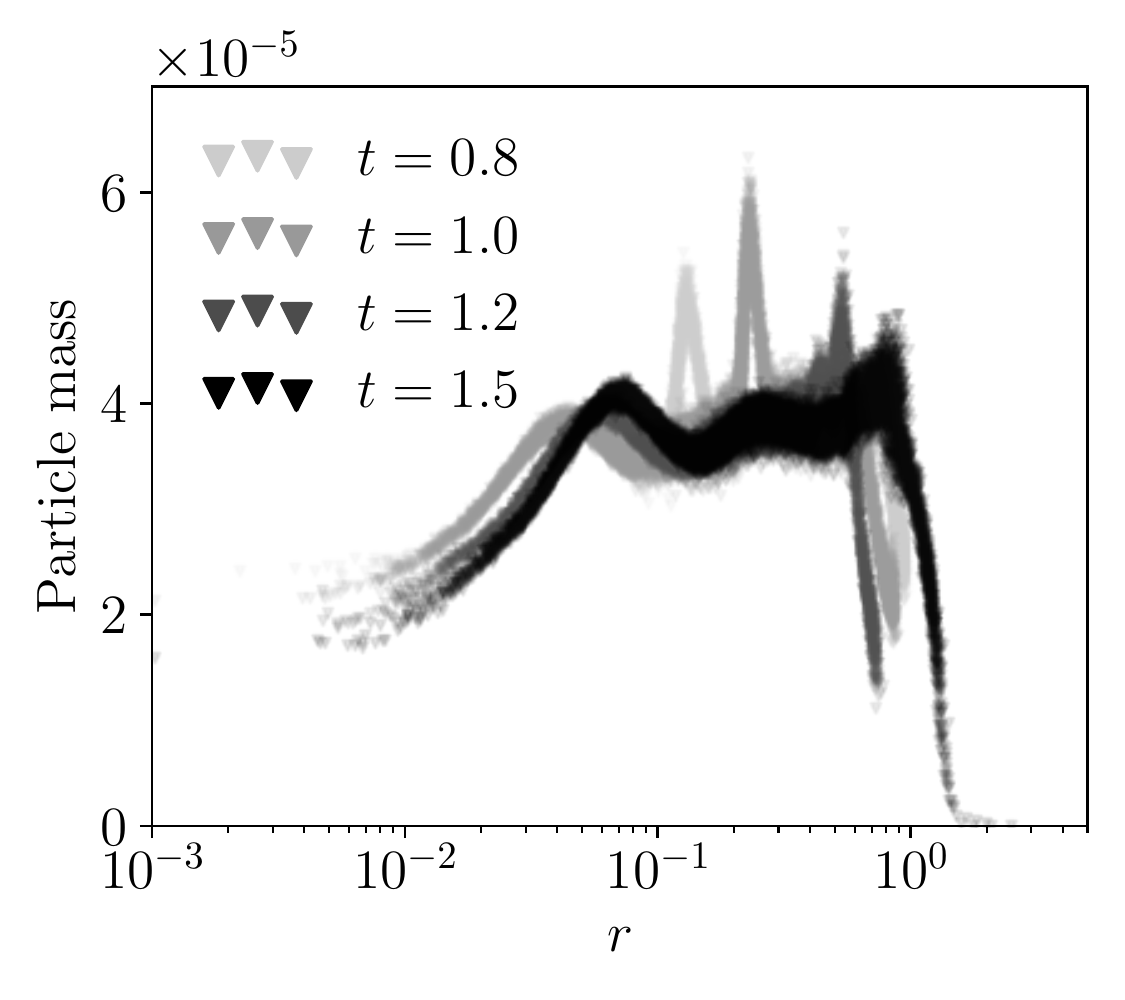}
   \caption{
       Evolution of the particle mass over time for the Evrard's spherical collapse using the MFV scheme.
       Once the shock reaches the outer region of the gaseous sphere, particles are accelerated into the vacuum.
       This causes a negative mass flux which drains these particles until their mass gets close to zero.
       This leads to unstable solutions of the hydrodynamics equations.
       This can be mitigated if particle merging is allowed or the gaseous sphere is embedded in a low density region rather than vacuum.
   \label{fig:evrard_3}}
\end{figure}

We have seen particle masses reaching extremely low values (zero or even negative due to round-off errors) in the MFV simulation.
In the Evrard collapse, this happens at late times, when the radially expanding shock moves beyond the gaseous sphere boundary, and external particles are accelerated into the vacuum.
Under such strong acceleration, these particles are totally devoid of their mass.
This is illustrated in figure~\ref{fig:evrard_3}, where the particle mass is shown at different simulation times as function of radial distance.

This is inherent to the setup of the problem, and can be mitigated, for example, by filling the remaining volume with low, constant density gas, and imposing periodic boundary conditions.
The shock would then naturally propagate through the edge of the sphere without expelling the outer particles.

\subsubsection{Zel'dovich pancake} \label{sec:zeldovich}

Another challenging test involving self-gravity, which also includes cosmological integration, is the \citet{Zeldovich1970} collapse.
The simulation starts with a uni-dimensional density perturbation in a Einstein-de Sitter universe,
which then collapses at redshift $z_c$, when a sheet is formed with diverging density.

This setup provides a strong test for cosmological codes.
The solution is highly anisotropic, whereas the schemes are built around the assumption that the particle distribution is near homogeneous and isotropic within a softening length.
Moreover, during the collapse, cold, gravitationally dominated flows develop, and the fluid can be artificially heated due to numerical errors and finite precision of the arithmetic operations (see section~\ref{sec:triple_energy}).
This test also provides a straightforward check of the cosmological integration in a comoving coordinate system (section~\ref{sec:comoving}).

\begin{figure*}
   \includegraphics[width=0.95\textwidth,trim=0mm 0mm -20mm 0mm]{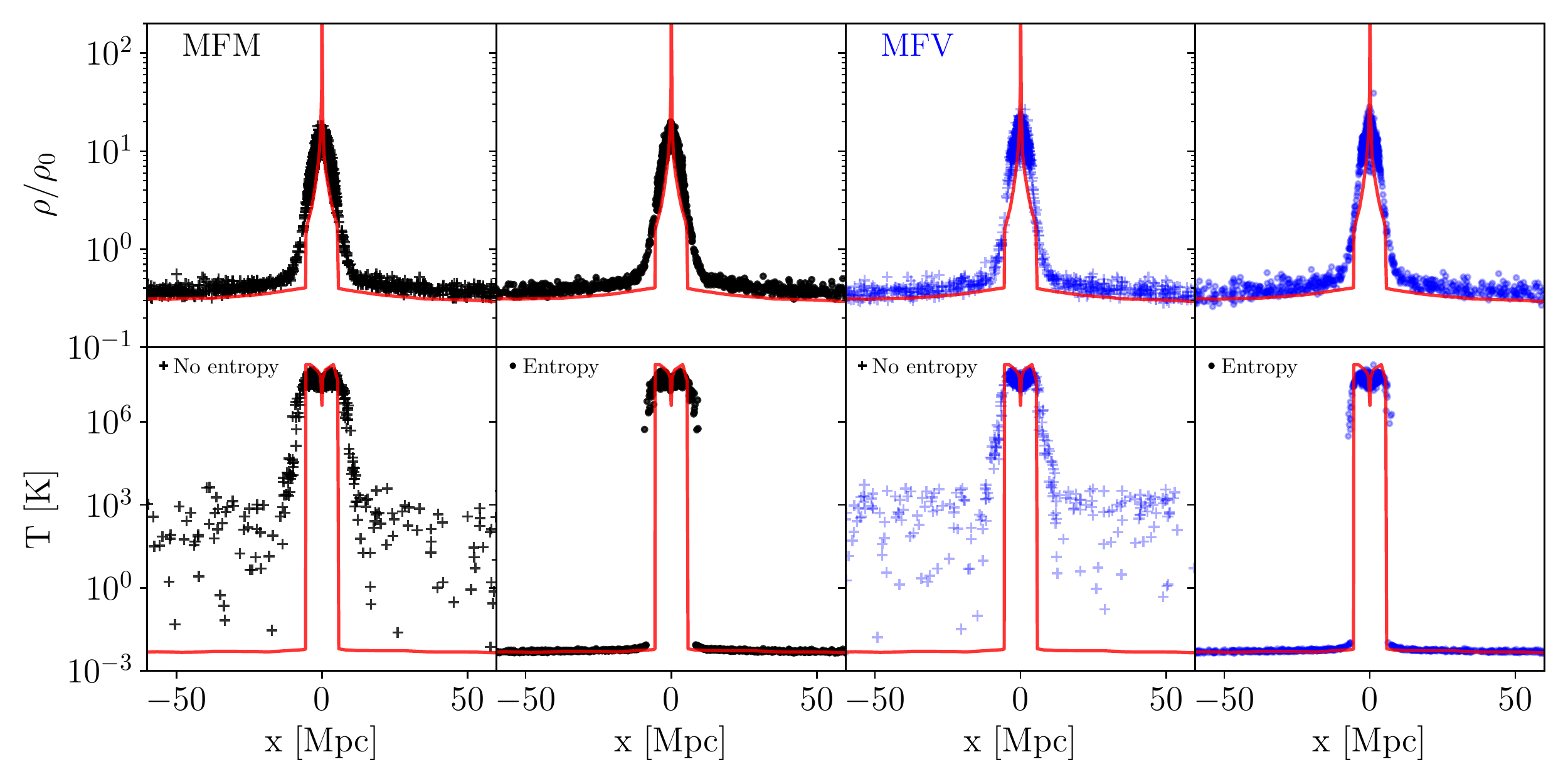}
   \caption{
      The Zel'dovich pancake test at $z=0$ (section~\ref{sec:zeldovich}).
      The MFM and MFV schemes are compared to the analytic solution in the left and right columns, respectively. Only $5\%$ of the particles are shown for clarity.
      The matter overdensity is plotted in the top row, whilst the gas temperature is plotted in the bottom row.
      We show the solution without entropy switch (cross symbols), where the internal energy is computed from $E_\textrm{tot}-E_\textrm{kin}$.
      Activating the entropy switch recovers the correct temperature of the background (dot symbols).
   \label{fig:zeldovich}}
\end{figure*}

The initial conditions are set up at $z_i=100$.
In a box of size $L=128~\Mpc$, $32^3$ gas particles are placed in a glass configuration using WVTICs \citep{Arth2019}.
Their position, velocity and temperature are perturbed as described in \citet{Zeldovich1970}
\begin{align}
   &x(q, z) = q - \frac{1+z_c}{1+z} \frac{\sin(kq)}{k}, \\
   &v_x(q,z) = -H_0 \frac{1+z_c}{\sqrt{1+z}} \frac{\sin(kq)}{k}, \\
   &T(q,z) = T_0 \left[ \left(\frac{1+z}{1+z_i}\right)^3 \frac{ \rho(q,z)}{\rho_c} \right]^{2/3},
   \label{eq:zeldo_pert}
\end{align}
being $q$ the unperturbed coordinate along the $x$-axis, such that the density is
\begin{equation}
    \rho (q,z) = \rho_c \left[ 1 - \frac{1+z_c}{1+z} \cos(kq) \right]^{-1}.
\end{equation}
Following \citet[e.g.,][]{Bryan1995, Bryan2014, Hopkins2015}, we set $z_c = 1$, $T_0=100$ K, $k=2\pi/L$, and $H_0 = 50~\mathrm{km}\,\mathrm{s}^{-1}\,\mathrm{Mpc}^{-1}$ .

The results at $z=0$ are shown in figure~\ref{fig:zeldovich}, for both the MFM (left panels) and MFV (right panels) schemes.
As consequence of the finite numerical precision, using internal or total energy in the integration of the equations of hydrodynamics leads to the wrong temperature evolution in smooth, cold flows.
This case is shown in figure~\ref{fig:zeldovich} as crosses.
It is evident that, although the density and velocity fields are correct, the temperature in the void is dominated by numerical noise.
This noise could, in turn, be amplified if another physics module uses the temperature as input variable.

When the entropy switch is used (denoted as points in figure~\ref{fig:zeldovich}), the temperature evolution in the infalling regions is correctly captured.
The switch is automatically disabled when the flow is no longer smooth, and thus the solution at the collapsed sheet is not changed.

\subsubsection{Galaxy cluster simulations} \label{sec:santabarbara}

The first simulation uses the initial conditions of the Santa Barbara cluster comparison project \citep{Frenk1999}.
This simulation has been extensively used for code comparison purposes, as it provides a controlled environment in which differences between numerical schemes can be studied.
The initial conditions are a cosmological box of size $L=64\Mpc$, and we chose the resolution of $2\times64^3$ particles.
This choice leads to a mass resolution of $m_\text{DM}=  3.1\times 10^{10} h^{-1} \Mo$ and $m_\text{gas}= 3.5 \times 10^9 h^{-1} \Mo$
and the following cosmological parameters were assumed: $H_0 = 50 \kmsMpc$, $\Omega=1$ and $\Omega_b=0.1$.

We successfully run this test using the MFM scheme, but we have encountered numerical issues in underdense regions when using MFV.
As illustrated in the Evrard collapse (section~\ref{sec:evrard}), the large accelerations generated by the shock produced by the collapse of the gas cloud can lead to numerical instabilities in the MFV scheme, as particles are depleted of their mass.
As opposed to \citet{Hopkins2015}, we have not implemented any particle merging procedure to mitigate this issue, as our future plans involve mainly the MFM scheme due to its Lagrangian nature.

\begin{figure}
   \includegraphics[width=\columnwidth,trim=5mm 10mm -10mm 0mm]{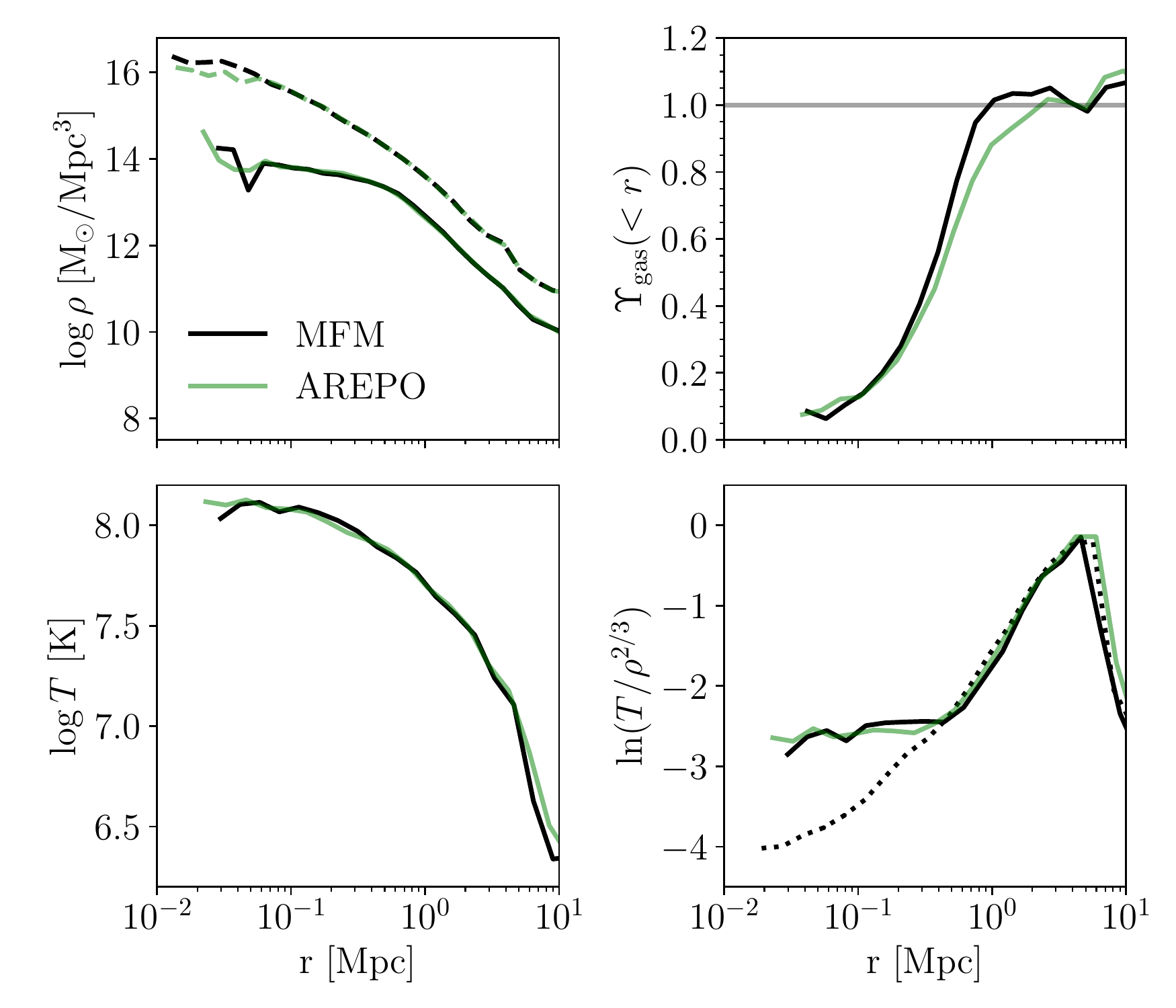}
   \caption{
      Radial profiles at $z=0$ of the Santa Barbara cluster comparison simulation for the MFM scheme (black) and \arepo (green).
      \textit{Upper left panel}. Density profile for baryons and dark matter, as continuous and dashed lines respectively.
      \textit{Upper right panel}. Cumulative gas fraction, $\Upsilon_\text{gas} (<r) = M_\text{gas}(<r)/M(<r)\, \Omega_m/\Omega_b$, measured from the center of the cluster. The cosmic baryon fraction is denoted as a gray line.
      \textit{Lower left panel}. Temperature profile.
      \textit{Lower right panel}. Entropy profile. The solution with a more restrictive gradient limiter is shown as a dotted line.
   \label{fig:santabarbara}}
\end{figure}

In figure~\ref{fig:santabarbara}, we show the solution for the MFM scheme in black, and a reference solution computed with the public version of \arepo \citep{Weinberger2020} in green.
For both runs, the same initial conditions were used.
Furthermore, the center of the cluster was determined using the same criteria, namely, the position of the dark matter particle with the lowest potential.

In accordance to previous results of the Santa Barbara cluster comparison performed with similar methods \citep[e.g.][]{Hopkins2015}, we find a cored entropy profile in our default configuration.
This is highly dependant on the particular choice of the gradient limiters and softening length \citep{Springel2010, Hopkins2015}.
To illustrate this, we show as a dotted line the entropy profile for a simulation with the more restrictive \citet{Barth89} limiter (see section~\ref{sec:limiter}).
In this case, the limiter completely suppresses artificial oscillations in the extrapolated variables, thus decreases entropy production.
Moreover, this limiter is more diffusive, and can excessively limit gradients in smooth flow regimes.
Hence, there is no clear correct numerical solution for the entropy profile.
We further note that our specific choice of entropy switch (section~\ref{sec:triple_energy}) does not have any impact on the inner entropy profile, but rather helps to recover the correct temperature evolution during the formation of the cluster.
At early times, the gas is cold and smooth, as it has not been confined within the potential well of dark matter haloes yet.
Therefore, the entropy switch can be used to accurately evolve the hydrodynamic state during the cluster formation.

The second simulation of the formation of a massive cluster employs the initial conditions of the nIFTy galaxy cluster comparison project \citep{Sembolini2016}.
The goal of the project was to provide initial conditions with recent cosmological parameters, and test modern codes beyond non-radiative simulations. Different models for galaxy formation have been compared in \citet{Sembolini2016b}, which is of special interest for future treating of the galaxy formation model to be implemented in \pkdgravthree.

The initial conditions were selected from the MultiDark cosmological simulation \citep{Prada2012}.
The cosmology is defined with parameters from the best-fit WMAP+BAO+SNI $\Lambda$CDM model \citep{Komatsu2011}: $\Omega_m = 0.27$, $\Omega_b=0.0469$, $\Omega_\Lambda=0.73$, $H_0 = 70 \kmsMpc$.
In \citet{Sembolini2016}, 12 different codes were used to make the comparison, running both collisionless and non-radiative simulations.
The mass resolution for the latter is $m_\text{DM}= 9.01 \times 10^8 h^{-1} \Mo$ and $m_\text{gas}= 1.9 \times 10^8 h^{-1} \Mo$.
Both have been performed with the version of \pkdgravthree presented in this work, and, following \citet{Sembolini2016}, the outputs were analysed with \ahf \citep{Knollmann2009} in order to identify subhalos and compute their radial density profiles.

\begin{figure}
   \includegraphics[width=\columnwidth,trim=5mm 10mm -10mm 0mm]{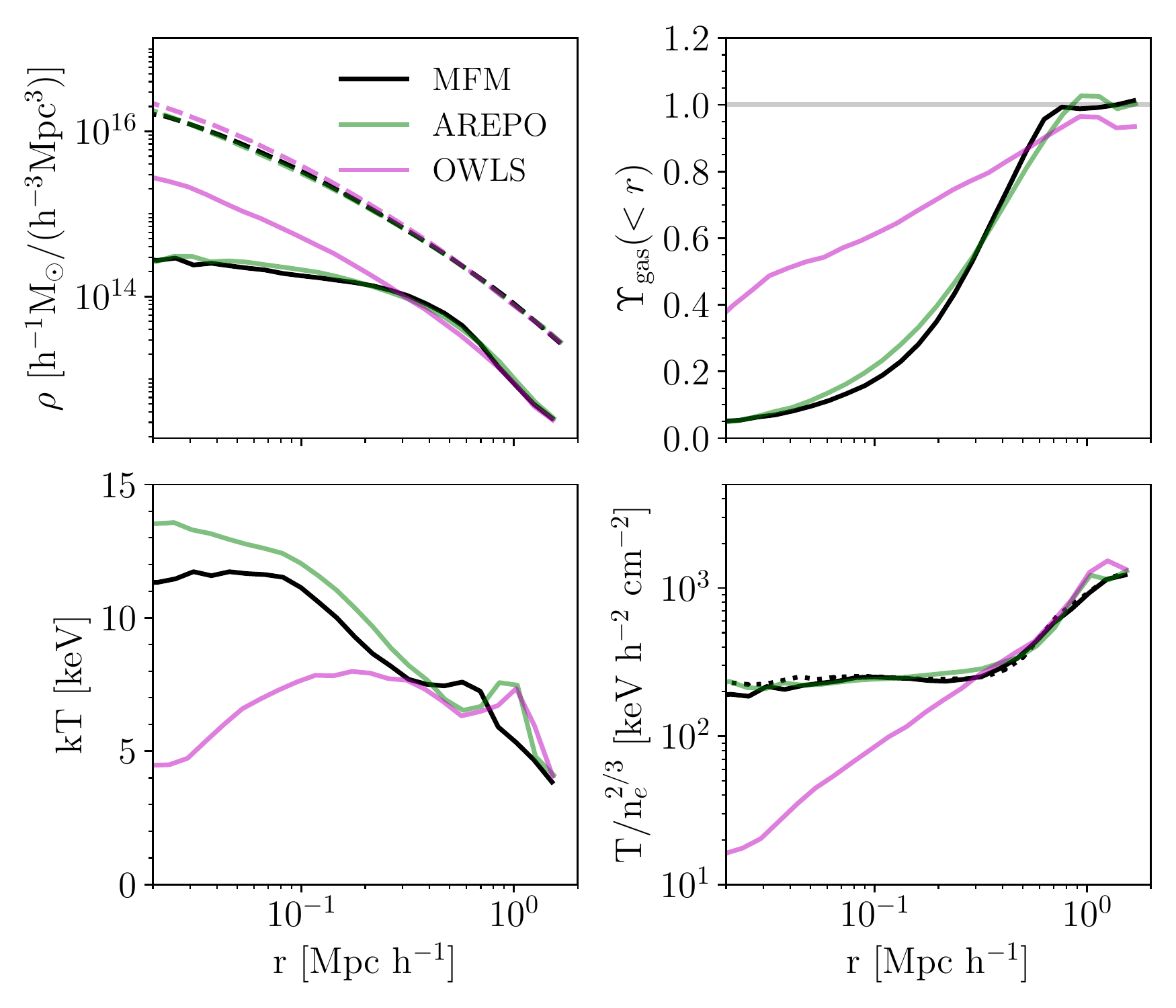}
   \caption{
      Radial profiles at $z=0$ of the nIFTy galaxy cluster comparison simulation for our MFM scheme (black) \arepo (green), and \textsc{gadget-owls} (magenta) as extracted from \ahf.
      \textit{Upper left panel}. Density profile for baryons and dark matter, as continuous and dashed lines respectively.
      \textit{Upper right panel}. Cumulative gas fraction, measured from the center of the cluster. The cosmic baryon fraction is denoted as a gray line.
      \textit{Lower left panel}. Temperature profile.
      \textit{Lower right panel}. Entropy profile. The solution with a more restrictive gradient limiter is shown as a dotted line for the MFM scheme.
   \label{fig:nifty}}
\end{figure}

We first checked that the code is giving the correct distribution of substructure and mass of the cluster by running the dark-matter only initial conditions (not shown here).
We then performed the non-radiative simulation. A summary of the profiles studied in \citet{Sembolini2016} is shown in figure~\ref{fig:nifty}, with the same panels of  figure~\ref{fig:santabarbara}.
To facilitate the comparison with nIFTY's results, we kept the scale of the axes of \citet{Sembolini2016}, and  the profiles were extracted processing the simulations output \ahf. We compare our results with those of two other codes, \textsc{gadget-owls} \citep{Schaye2010} and \arepo, from \citet{Sembolini2016}.
These were chosen as a representation of the extremes for the inner region entropy profiles.

As for the Santa Barbara test, the MFM scheme yields results similar to \arepo for all profiles.
In \citet{Sembolini2016} \citep[and][]{Frenk1999} codes are classified according to the shape of the inner entropy profile, whether an entropy core is formed or not.
Consistent with the results shown for the Santa Barbara cluster, the MFM method produces an entropy core.
However, contrary to the Santa Barbara test, when using the more restrictive limiter (dotted line), the change of the inner entropy profile is negligible.
This suggests that the resolution may have an important role in the sensitivity to the change of the slope limiter: the higher the resolution, the smaller the discrepancy introduced by the slope limiter.

\subsection{Scaling tests} \label{sec:scaling}

We tested the code for weak and strong scaling with cosmological, hydrodynamic initial conditions.
The tests were performed on the multicore partition of Piz Daint, which hosts two Intel E5-2695v4 CPUs per node (36 cores per node).
Each node has 64 GB of RAM memory.

The size of the cosmological volume is kept fixed for the strong scaling text, whereas, for weak scaling, it increases proportionally to the number of nodes used for the test:
\begin{equation}
    L = \frac{N}{2} h^{-1} \Mpc, \label{eq:N_scaling}
\end{equation}
where $L$ is the length of the side of the cubic volume, $N\simeq n^{1/3}N_0$ is the linear number of particles and $n$ is the number of nodes.
In order to fill the memory of one node, we set $N_0=370$. The total number of particles is then $N_\text{tot}=2\times N^3$.
The number of particles per core is $\simeq 2.8\times 10^6$, corresponding to $\simeq 1.5~\mathrm{GB}$ per core.
We used the following cosmological parameters: $H_0 = 67.7 \kmsMpc$, $\Omega_m = 0.32$, $\Omega_b = 0.048$, $\Omega_\Lambda = 0.68$, $\sigma_8 = 0.83$.
This gives the particle masses, $m_\text{DM}=9 \times 10^9 h^{-1} \Mo$ and $m_\text{gas}=1.6 \times 10^9 h^{-1} \Mo$, for dark matter and baryons, respectively.
The initial conditions were generated at $z=49$.
We used one MPI task per socket, and one thread per physical core.
The timing was obtained by averaging over the three first full steps.

The use of high redshift initial conditions eases the scaling of the code, because the homogeneity of the distribution of particles gives a shallower tree.
Moreover, the actual load balancing scheme of \pkdgravthree simply partitions the domain such that all cores have an equal number of particles.
This was designed for large-scale cosmological simulations, and is not well suited for highly clustered particle distributions and zoom-in simulations.
In the future, the load balancing scheme will be upgraded to improve its performance.

\begin{figure}
   \centering
   \includegraphics[width=\columnwidth,trim=0mm 10mm -10mm 0mm]{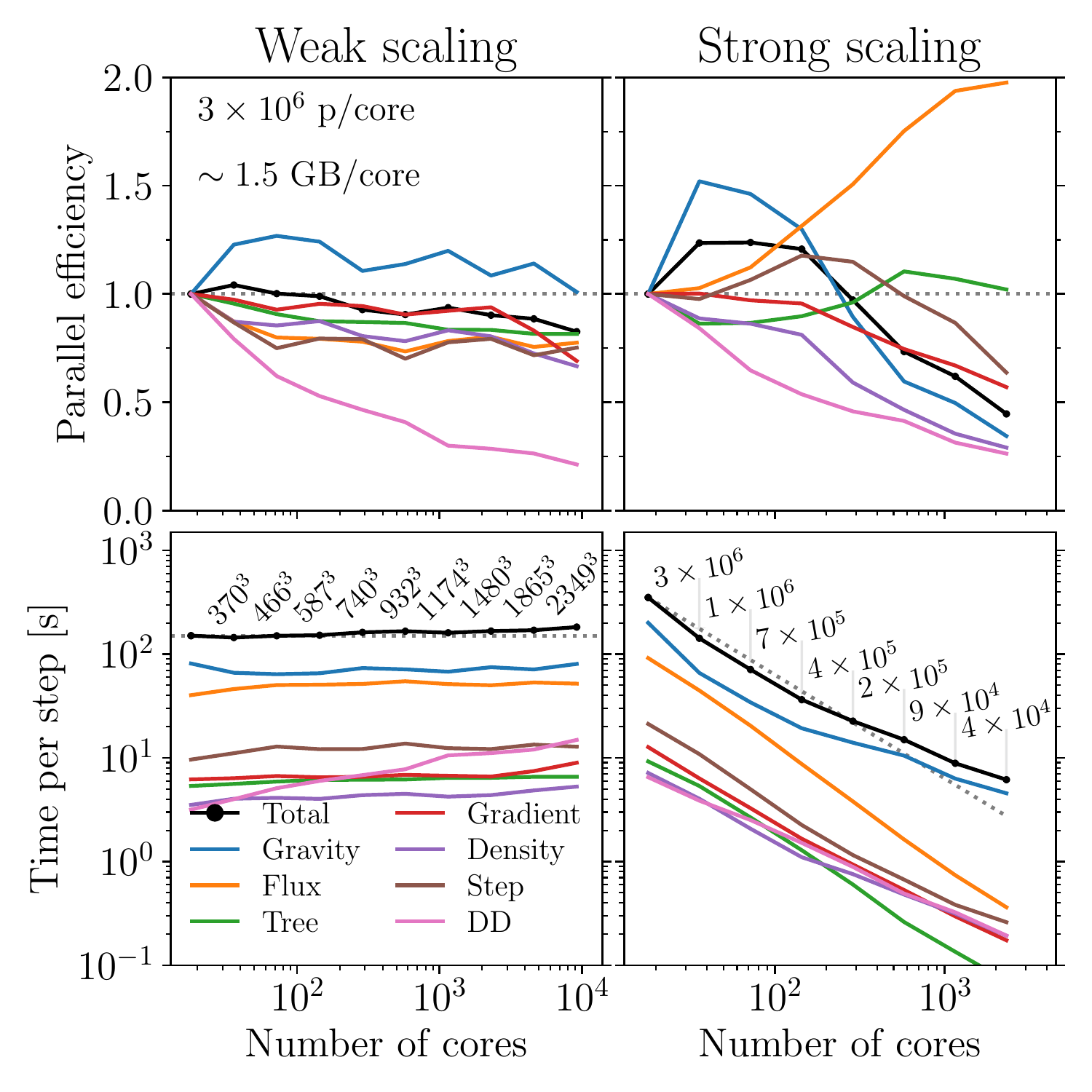}
   \caption{Scaling test on the multicore partition of Piz Daint in the high redshift regime.
       The parallel efficiency and the average time per step are shown in the top and bottom rows for the weak (\textit{left column}) and strong (\textit{right column}) scaling.
       For the weak scaling, the particle load per core is indicated in the upper panel.
       In the lower panel, the labels indicate the problem size, $N$ for the weak scaling, and particles per core for the strong scaling.
       \pkdgravthree shows very good weak scaling up to $10^4$ cores, and good strong scaling down to tens of thousands of particles per core.
       The hydrodynamic solver described in this work shows good scaling in both tests.
       \label{fig:daint_scaling}
   }
\end{figure}

We show the results of the weak (left column) and strong (right column) scaling tests in figure~\ref{fig:daint_scaling}.
Efficiency and the average time per step are in the top and bottom row, respectively.
For each weak scaling simulation, we indicate the value of $N$ (bottom-left panel). For each strong scaling simulation, we indicate the number of particles per core (bottom-right panel).

The weak scaling test shows an efficiency of about $80$ percent on 256 nodes (9216 cores).
Gravity and flux computations take the longest time per step, as expected.
We note that the Intel E5-2695v4 CPU does not support the AVX-512 instruction set. Vectorization can speed up the flux and gravity computation, according to our tests, by up to a factor of two.
The density and gradients calculations, that rely on the neighbours search, show both good performance and scaling.
As it is usually found \citep[e.g.,][]{Springel2021}, the domain decomposition (DD) cannot scale optimally, as the workload increases with the number of MPI processes.
In this test, the time spent in the domain decomposition is a factor of four larger going from one (2 MPI tasks) to 256 (512 MPI tasks) nodes.

The scaling starts to be suboptimal around tens of thousands of particles per core in the strong scaling test.
The most remarkable feature is the superlinear scaling of the flux computation.
This is the result of more efficient cache usage: as the particle load per core decreases, a larger fraction of the data can be stored in the CPU cache.
If the problem is significantly memory-bounded, this can yield a speed-up proportional to the increase in memory bandwidth.
Superlinear behaviour is also seen for the gravity and density calculation and for the tree construction for different ranges of node numbers.

The scaling can be compared to that of the N-body version of \pkdgravthree on Piz Daint, presented in \citet{Potter2017}.\footnote{They use the 'old' GPU partition of Piz Daint consisting in nodes with one Intel\textregistered\ E5-2670 CPU, one Nvidia K20X and 32 GB of memory.}
The weak scaling test was performed with $4.7 \times 10^8$ particles per node ($5.9 \times 10^7$ per core).
This figure is almost an order of magnitude higher than the particle per core of our version of the code implementing the hydrodynamics.
They show almost perfect scaling up to about $\sim 8\times 10^4$ cores.

Their strong scaling test reaches 50 percent efficiency for the smallest simulation on 600 nodes (4800 cores).
Our reaches similar efficiencies on 64 nodes (2304 cores).
However, the number of particle for our strong scaling tests is significantly lower than that of \citet{Potter2017} ($2\times 370^3$ vs. $1000^3$).
If we compare the particles per core, the efficiency drops below 50 percent at $\sim 2\times 10^5$ particles/core, whereas, in our case, this occurs at $4 \times 10^4$ particles/core, implying that the version of \pkdgravthree in this work may scale better than that of \citet{Potter2017}.
However, we shall remind the reader that:
the code has evolved and improved significantly since 2017;
the architectures on which the tests have been performed are different;
the number of cores per node is much larger in the current supercomputer, which considerably reduces the communication over the network;
the tests presented here do not use GPU acceleration.

Compared with the N-body weak scaling results of \gadgetfour \citep[figure 62 of][]{Springel2021}, both the tree and gravity computations show substantially better scaling in \pkdgravthree.
It must be noted, however, that we only reach $N=2349$ in Piz Daint, whereas they got up to $N=6656$ in the gravity-only configuration.
The load in their case is of $3 \times 10^6$ particles per core, comparable to that of our scaling tests.
The rate of gravity calculations expressed in particles per second per core is $2.9 \times 10^4$ for \gadgetfour and $3.5 \times 10^4$ for \pkdgravthree.
The system used for the scaling test of \gadgetfour had more cores per nodes and supported the AVX-512 instructions set, which would speed-up both the gravity and the flux computation in \pkdgravthree, and yield better performance than \gadgetfour.
Regarding the strong scaling test \citep[figure 63 of][]{Springel2021}, scaling is largely degraded by the domain decomposition, which takes most of the computational time when more than $10^4$ cores are used ($\sim 13000$ particles per core), even at high redshift.
In the scaling test including baryons, the efficiency of the hydrodynamic solver (an SPH scheme) degrade faster than that of gravity.
However, it never dominates the computing time due to the relative simplicity of the SPH scheme compared to our hydrodynamic scheme.
At large enough core numbers, the domain decomposition becomes the bottleneck.

Our scaling tests can also be compared to those performed with \swift and presented in \citet{Borrow2018},
where the tests are performed with simulations at $z=0.1$, when matter is highly clustered.
They performed the strong scaling test up to 512 threads.
The total number of particles in the initial conditions is $6 \times 10^7$, which, for 512 threads, gives the average particle load per core of $1.1 \times 10^6$.
The scaling departs from linearity around 16 threads, and flatten out for more than 64 threads, although their initial conditions contains twenty times more particles than ours.
However, it is expected that the clustering of particles at low redshift degrades the performance of the code.
They also show weak scaling results with a particle load of $7 \times 10^6$ particles per core,
maintaining an efficiency of 80 percent with 4096 threads.
This is similar to the result of the weak scaling test we showed above, with the latter performed at high redshift.
The scaling tests in \citet{Borrow2018} do not detail the performance of different parts of the code, and a more detailed comparison with our results is not possible.

%%%%%%%%%%

\section{Conclusions} \label{sec:Conclusions}

We have implemented two mesh-less hydrodynamic methods into the state-of-the-art N-body code \pkdgravthree, providing the code with the capability of performing cosmological hydrodynamic simulations.
This work aims to be the first in a series of \pkdgravthree upgrades towards a fully consistent code for studies of galaxy formation and evolution and can be summarized as follows:
\begin{enumerate}
   \item The MFM and MFV hydrodynamic schemes \citep{Lanson2008a, Gaburov2011, Hopkins2015} have been implemented within  \pkdgravthree (section~\ref{sec:Hydrodynamics}), taking advantage of the low-level structure of the code.
         The implementation is largely based on the original work of \citet{Lanson2008a} and \citet{Gaburov2011}, and provides an independent test of the feasibility of the scheme.
         Hydrodynamics has been coupled to gravity, and the code can perform both hydrodynamic simulations with self-gravity and cosmological hydrodynamic simulations.
   \item An extensive suite of numerical tests has been used to check the correctness of the implementation. We proved that the schemes
         are second order accurate (figure~\ref{fig:soundwaves}),
         conserve energy to machine accuracy (figure~\ref{fig:sedov}),
         and give accurate results when including self-gravity in a cosmological context (figure~\ref{fig:santabarbara}).
   \item We extended the original code by adding support for multiple particle types, and improved it by
         optimising the neighbours search to reduce communication,
         adding vectorization of the computationally expensive hydrodynamic loop,
         and adopting the HDF5 file format for storing output data.
   \item We have shown that, in its current state, the code can scale to $10^4$ cores (figure~\ref{fig:daint_scaling}), maintaining a parallel efficiency above $80$ percent in the weak scaling test.
         \pkdgravthree gives good performance in strong scaling tests, down to a workload of tens of thousands of particles per core.
\end{enumerate}

The code presented in this work will be the framework for the implementation of physics modules for cosmological simulations of galaxy formation in large volumes.
This will be described in a future article.
In addition, its performance will be further improved and tuned in the future.
The code and all the tests performed in this work have been publicly released.

\section*{Data availability}

The code used for this research, documentation, test cases and analysis scripts are public and can be accessed from the project webpage: \url{https://research.iac.es/proyecto/PKDGRAV3}.
There, the tagged versions used in this work can be accessed.
Access to the latest stable development version can be requested to the authors.

\section*{Acknowledgements}

We thank Andrea Negri and Alex Massaro Ach\'a for helping in the revision of the final draft. We are grateful to Alexander Knebe for useful discussions about the nIFTy cluster comparison and for providing support for \ahf.
IAA and CDV are supported by the Spanish Ministry of Science and Innovation (MICIU/FEDER) through research grant PGC2018-094975-C22 and PID2021-122603NB-C22.
CDV acknowledges support from MICIU through grant RYC-2015-18078.
This research made use of computing time on the high-performance computing system Deimos/Diva of the \textit{Instituto de Astrof\'isica de Canarias}.
The author thankfully acknowledges the technical expertise and assistance provided by the Spanish Supercomputing Network (\textit{Red Española de Supercomputación}), as well as the computer resources used: the LaPalma supercomputer, located at the \textit{Instituto de Astrof\'isica de Canarias}, and MareNostrum4 and MinoTauro at the Barcelona Supercomputing Center (RES-AECT-2020-2-0003).
We acknowledge the access to Piz Daint at the Swiss National Supercomputing Centre, Switzerland under the University of Zurich's share with the project ID UZH4.
The following Python packages have been extensively used for this research:
\textsc{numpy}\footnote{\url{https://numpy.org}} \citep{Harris2020},
\textsc{matplotlib}\footnote{\url{https://matplotlib.org/}} \citep{Hunter2007} and
\textsc{pynbody}\footnote{\url{https://pynbody.github.io/pynbody/}} \citep{pynbody}.

\bibliographystyle{mnras}
\bibliography{biblio}

\begin{thebibliography}{}
\makeatletter
\relax
\def\mn@urlcharsother{\let\do\@makeother \do\$\do\&\do\#\do\^\do\_\do\%\do\~}
\def\mn@doi{\begingroup\mn@urlcharsother \@ifnextchar [ {\mn@doi@}
  {\mn@doi@[]}}
\def\mn@doi@[#1]#2{\def\@tempa{#1}\ifx\@tempa\@empty \href
  {http://dx.doi.org/#2} {doi:#2}\else \href {http://dx.doi.org/#2} {#1}\fi
  \endgroup}
\def\mn@eprint#1#2{\mn@eprint@#1:#2::\@nil}
\def\mn@eprint@arXiv#1{\href {http://arxiv.org/abs/#1} {{\tt arXiv:#1}}}
\def\mn@eprint@dblp#1{\href {http://dblp.uni-trier.de/rec/bibtex/#1.xml}
  {dblp:#1}}
\def\mn@eprint@#1:#2:#3:#4\@nil{\def\@tempa {#1}\def\@tempb {#2}\def\@tempc
  {#3}\ifx \@tempc \@empty \let \@tempc \@tempb \let \@tempb \@tempa \fi \ifx
  \@tempb \@empty \def\@tempb {arXiv}\fi \@ifundefined
  {mn@eprint@\@tempb}{\@tempb:\@tempc}{\expandafter \expandafter \csname
  mn@eprint@\@tempb\endcsname \expandafter{\@tempc}}}

\bibitem[\protect\citeauthoryear{Agertz et~al.,}{Agertz
  et~al.}{2007}]{Agertz2007}
Agertz O.,  et~al., 2007, \mn@doi [\mnras] {10.1111/j.1365-2966.2007.12183.x},
  380, 963

\bibitem[\protect\citeauthoryear{Arth, Donnert, Steinwandel, B{\"{o}}ss,
  Halbesma, P{\"{u}}tz, Hubber  \& Dolag}{Arth et~al.}{2019}]{Arth2019}
Arth A.,  Donnert J.,  Steinwandel U.,  B{\"{o}}ss L.,  Halbesma T.,
  P{\"{u}}tz M.,  Hubber D.,   Dolag K.,  2019, arXiv

\bibitem[\protect\citeauthoryear{Barnes et~al.,}{Barnes
  et~al.}{2017}]{Barnes2017}
Barnes D.~J.,  et~al., 2017, \mn@doi [\mnras] {10.1093/MNRAS/STX1647}, 471,
  1088

\bibitem[\protect\citeauthoryear{Barth \& Jespersen}{Barth \&
  Jespersen}{1989}]{Barth89}
Barth T.,  Jespersen D.,  1989, in 27th Aerospace Sciences Meeting. American
  Institute of Aeronautics and Astronautics, Reston, Virigina,
  \mn@doi{10.2514/6.1989-366}, \url
  {http://arc.aiaa.org/doi/10.2514/6.1989-366}

\bibitem[\protect\citeauthoryear{Borrow, Bower, Draper, Gonnet  \&
  Schaller}{Borrow et~al.}{2018}]{Borrow2018}
Borrow J.,  Bower R.~G.,  Draper P.~W.,  Gonnet P.,   Schaller M.,  2018,
  arXiv:1807.01341 [astro-ph]

\bibitem[\protect\citeauthoryear{Borrow, Schaller, Bower  \& Schaye}{Borrow
  et~al.}{2022}]{Borrow2022}
Borrow J.,  Schaller M.,  Bower R.~G.,   Schaye J.,  2022, \mn@doi [\mnras]
  {10.1093/mnras/stab3166}, 511, 2367

\bibitem[\protect\citeauthoryear{Braspenning, Schaye, Borrow  \&
  Schaller}{Braspenning et~al.}{2022}]{Braspenning2022}
Braspenning J.,  Schaye J.,  Borrow J.,   Schaller M.,  2022, arXiv

\bibitem[\protect\citeauthoryear{Bryan, Norman, Stone, Cen  \& Ostriker}{Bryan
  et~al.}{1995}]{Bryan1995}
Bryan G.~L.,  Norman M.~L.,  Stone J.~M.,  Cen R.,   Ostriker J.~P.,  1995,
  \mn@doi [Computer Physics Communications] {10.1016/0010-4655(94)00191-4}, 89,
  149

\bibitem[\protect\citeauthoryear{Bryan et~al.,}{Bryan et~al.}{2014}]{Bryan2014}
Bryan G.~L.,  et~al., 2014, \mn@doi [\apjs] {10.1088/0067-0049/211/2/19}, 211

\bibitem[\protect\citeauthoryear{Crain et~al.,}{Crain et~al.}{2015}]{Crain2015}
Crain R.~A.,  et~al., 2015, \mn@doi [\mnras] {10.1093/mnras/stv725}, 450, 1937

\bibitem[\protect\citeauthoryear{Cullen \& Dehnen}{Cullen \&
  Dehnen}{2010}]{Cullen2010}
Cullen L.,  Dehnen W.,  2010, \mn@doi [\mnras]
  {10.1111/j.1365-2966.2010.17158.x}, 408, 669

\bibitem[\protect\citeauthoryear{Dav{\'{e}}, Thompson  \& Hopkins}{Dav{\'{e}}
  et~al.}{2016}]{Dave2016}
Dav{\'{e}} R.,  Thompson R.,   Hopkins P.~F.,  2016, \mn@doi [\mnras]
  {10.1093/mnras/stw1862}, 462, 3265

\bibitem[\protect\citeauthoryear{Dav{\'{e}}, Angl{\'{e}}s-Alc{\'{a}}zar,
  Narayanan, Li, Rafieferantsoa  \& Appleby}{Dav{\'{e}}
  et~al.}{2019}]{Dave2019}
Dav{\'{e}} R.,  Angl{\'{e}}s-Alc{\'{a}}zar D.,  Narayanan D.,  Li Q.,
  Rafieferantsoa M.~H.,   Appleby S.,  2019, \mn@doi [\mnras]
  {10.1093/mnras/stz937}, 486, 2827

\bibitem[\protect\citeauthoryear{Dehnen \& Aly}{Dehnen \&
  Aly}{2012}]{Dehnen2012}
Dehnen W.,  Aly H.,  2012, \mn@doi [\mnras] {10.1111/j.1365-2966.2012.21439.x},
  425, 1068

\bibitem[\protect\citeauthoryear{Deilmann, Kuah, Corden  \& Sabah}{Deilmann
  et~al.}{2012}]{Intel2012}
Deilmann M.,  Kuah K.,  Corden M.,   Sabah M.,  2012, Technical report, {A
  Guide to Vectorization with Intel {\textregistered} C ++ Compilers}, \url
  {https://software.intel.com/sites/default/files/m/4/8/8/2/a/31848-CompilerAutovectorizationGuide.pdf}.
Intel Coorporation, \url
  {https://software.intel.com/sites/default/files/m/4/8/8/2/a/31848-CompilerAutovectorizationGuide.pdf}

\bibitem[\protect\citeauthoryear{Diemand, Moore  \& Stadel}{Diemand
  et~al.}{2004}]{Diemand2004}
Diemand J.,  Moore B.,   Stadel J.,  2004, \mn@doi [\mnras]
  {10.1111/j.1365-2966.2004.08094.x}, 353, 624

\bibitem[\protect\citeauthoryear{Dubois, Gavazzi, Peirani  \& Silk}{Dubois
  et~al.}{2013}]{Dubois2013}
Dubois Y.,  Gavazzi R.,  Peirani S.,   Silk J.,  2013, \mn@doi [\mnras]
  {10.1093/mnras/stt997}, 433, 3297

\bibitem[\protect\citeauthoryear{Dubois et~al.,}{Dubois
  et~al.}{2014}]{Dubois2014}
Dubois Y.,  et~al., 2014, \mn@doi [\mnras] {10.1093/mnras/stu1227}, 444, 1453

\bibitem[\protect\citeauthoryear{Evrard}{Evrard}{1988}]{Evrard1988}
Evrard A.~E.,  1988, \mn@doi [\mnras] {10.1093/mnras/235.3.911}, 235, 911

\bibitem[\protect\citeauthoryear{Frenk et~al.,}{Frenk et~al.}{1999}]{Frenk1999}
Frenk C.~S.,  et~al., 1999, \mn@doi [\apj] {10.1086/307908}, 525, 554

\bibitem[\protect\citeauthoryear{Gaburov \& Nitadori}{Gaburov \&
  Nitadori}{2011}]{Gaburov2011}
Gaburov E.,  Nitadori K.,  2011, \mn@doi [\mnras]
  {10.1111/j.1365-2966.2011.18313.x}, 414, 129

\bibitem[\protect\citeauthoryear{Garrison, Eisenstein  \& Pinto}{Garrison
  et~al.}{2019}]{Garrison2019}
Garrison L.~H.,  Eisenstein D.~J.,   Pinto P.~A.,  2019, \mn@doi [\mnras]
  {10.1093/mnras/stz634}, 485, 3370

\bibitem[\protect\citeauthoryear{Gingold \& Monaghan}{Gingold \&
  Monaghan}{1977}]{Gingold77}
Gingold R.~A.,  Monaghan J.~J.,  1977, \mn@doi [\mnras]
  {10.1093/mnras/181.3.375}, 181, 375

\bibitem[\protect\citeauthoryear{Greengard \& Rokhlin}{Greengard \&
  Rokhlin}{1997}]{Greengard1997}
Greengard L.,  Rokhlin V.,  1997, \mn@doi [Journal of Computational Physics]
  {10.1006/jcph.1997.5706}, 135, 280

\bibitem[\protect\citeauthoryear{Gresho \& Chan}{Gresho \&
  Chan}{1990}]{Gresho1990}
Gresho P.~M.,  Chan S.~T.,  1990, \mn@doi [International Journal for Numerical
  Methods in Fluids] {10.1002/fld.1650110510}, 11, 621

\bibitem[\protect\citeauthoryear{Harris et~al.,}{Harris
  et~al.}{2020}]{Harris2020}
Harris C.~R.,  et~al., 2020, \mn@doi [\nat] {10.1038/s41586-020-2649-2}, 585,
  357

\bibitem[\protect\citeauthoryear{Hernquist, Bouchet  \& Suto}{Hernquist
  et~al.}{1991}]{Hernquist1991}
Hernquist L.,  Bouchet F.~R.,   Suto Y.,  1991, \mn@doi [\apjs]
  {10.1086/191530}, 75, 231

\bibitem[\protect\citeauthoryear{Hopkins}{Hopkins}{2015}]{Hopkins2015}
Hopkins P.~F.,  2015, \mn@doi [\mnras] {10.1093/mnras/stv195}, 450, 53

\bibitem[\protect\citeauthoryear{Hu, Naab, Walch, Moster  \& Oser}{Hu
  et~al.}{2014}]{Hu2014}
Hu C.~Y.,  Naab T.,  Walch S.,  Moster B.~P.,   Oser L.,  2014, \mn@doi
  [\mnras] {10.1093/mnras/stu1187}, 443, 1173

\bibitem[\protect\citeauthoryear{Hubber, Rosotti  \& Booth}{Hubber
  et~al.}{2018}]{Hubber2018}
Hubber D.~A.,  Rosotti G.~P.,   Booth R.~A.,  2018, \mn@doi [\mnras]
  {10.1093/mnras/stx2405}, 473, 1603

\bibitem[\protect\citeauthoryear{Hunter}{Hunter}{2007}]{Hunter2007}
Hunter J.~D.,  2007, \mn@doi [Computing in Science & Engineering]
  {10.1109/MCSE.2007.55}, 9, 90

\bibitem[\protect\citeauthoryear{Knabenhans et~al.,}{Knabenhans
  et~al.}{2019}]{Euclid2019}
Knabenhans M.,  et~al., 2019, \mn@doi [\mnras] {10.1093/mnras/stz197}, 484,
  5509

\bibitem[\protect\citeauthoryear{Knollmann \& Knebe}{Knollmann \&
  Knebe}{2009}]{Knollmann2009}
Knollmann S.~R.,  Knebe A.,  2009, \mn@doi [\apjs]
  {10.1088/0067-0049/182/2/608}, 182, 608

\bibitem[\protect\citeauthoryear{Komatsu et~al.,}{Komatsu
  et~al.}{2011}]{Komatsu2011}
Komatsu E.,  et~al., 2011, \mn@doi [\apjs] {10.1088/0067-0049/192/2/18}, 192,
  18

\bibitem[\protect\citeauthoryear{Lanson \& Vila}{Lanson \&
  Vila}{2008a}]{Lanson2008a}
Lanson N.,  Vila J.-P.,  2008a, \mn@doi [SIAM Journal on Numerical Analysis]
  {10.1137/S0036142903427718}, 46, 1912

\bibitem[\protect\citeauthoryear{Lanson \& Vila}{Lanson \&
  Vila}{2008b}]{Lanson2008b}
Lanson N.,  Vila J.-P.,  2008b, \mn@doi [SIAM Journal on Numerical Analysis]
  {10.1137/S003614290444739X}, 46, 1935

\bibitem[\protect\citeauthoryear{LeVeque}{LeVeque}{1992}]{LeVeque1992}
LeVeque R.~J.,  1992, {Numerical Methods for Conservation Laws}.
Birkh{\"{a}}user Basel, Basel, \mn@doi{10.1007/978-3-0348-8629-1}, \url
  {http://link.springer.com/10.1007/978-3-0348-8629-1}

\bibitem[\protect\citeauthoryear{Leinhardt \& Stewart}{Leinhardt \&
  Stewart}{2009}]{Leinhardt2009}
Leinhardt Z.~M.,  Stewart S.~T.,  2009, \mn@doi [\icarus]
  {10.1016/j.icarus.2008.09.013}, 199, 542

\bibitem[\protect\citeauthoryear{Leinhardt, Richardson  \& Quinn}{Leinhardt
  et~al.}{2000}]{Leinhardt2000}
Leinhardt Z.~M.,  Richardson D.~C.,   Quinn T.,  2000, \mn@doi [\icarus]
  {10.1006/icar.2000.6370}, 146, 133

\bibitem[\protect\citeauthoryear{Lucy}{Lucy}{1977}]{Lucy77}
Lucy L.~B.,  1977, \mn@doi [\aj] {10.1086/112164}, 82, 1013

\bibitem[\protect\citeauthoryear{McCarthy, Schaye, Bird  \& {Le Brun}}{McCarthy
  et~al.}{2017}]{McCarthy2017}
McCarthy I.~G.,  Schaye J.,  Bird S.,   {Le Brun} A. M.~C.,  2017, \mn@doi
  [\mnras] {10.1093/mnras/stw2792}, 465, 2936

\bibitem[\protect\citeauthoryear{Menon, Wesolowski, Zheng, Jetley, Kale, Quinn
  \& Governato}{Menon et~al.}{2015}]{Menon2015}
Menon H.,  Wesolowski L.,  Zheng G.,  Jetley P.,  Kale L.,  Quinn T.,
  Governato F.,  2015, \mn@doi [Computational Astrophysics and Cosmology]
  {10.1186/s40668-015-0007-9}, 2, 1

\bibitem[\protect\citeauthoryear{Morton, Khochfar  \& Wu}{Morton
  et~al.}{2022}]{Morton2022}
Morton B.,  Khochfar S.,   Wu Z.,  2022, arXiv

\bibitem[\protect\citeauthoryear{Nesvorn{\'{y}}, Youdin  \&
  Richardson}{Nesvorn{\'{y}} et~al.}{2010}]{Nesvorny2010}
Nesvorn{\'{y}} D.,  Youdin A.~N.,   Richardson D.~C.,  2010, \mn@doi [\aj]
  {10.1088/0004-6256/140/3/785}, 140, 785

\bibitem[\protect\citeauthoryear{Pillepich et~al.,}{Pillepich
  et~al.}{2018}]{Pillepich2018}
Pillepich A.,  et~al., 2018, \mn@doi [\mnras] {10.1093/mnras/stx3112}, 475, 648

\bibitem[\protect\citeauthoryear{Pontzen, Ro{\v{s}}kar, Stinson  \&
  Woods}{Pontzen et~al.}{2013}]{pynbody}
Pontzen A.,  Ro{\v{s}}kar R.,  Stinson G.,   Woods R.,  2013, {pynbody:
  N-Body/SPH analysis for python}

\bibitem[\protect\citeauthoryear{Potter, Stadel  \& Teyssier}{Potter
  et~al.}{2017}]{Potter2017}
Potter D.,  Stadel J.,   Teyssier R.,  2017, \mn@doi [Computational
  Astrophysics and Cosmology] {10.1186/s40668-017-0021-1}, 4, 2

\bibitem[\protect\citeauthoryear{Power, Navarro, Jenkins, Frenk, White,
  Springel, Stadel  \& Quinn}{Power et~al.}{2003}]{Power2003}
Power C.,  Navarro J.~F.,  Jenkins A.,  Frenk C.~S.,  White S. D.~M.,  Springel
  V.,  Stadel J.,   Quinn T.,  2003, \mn@doi [\mnras]
  {10.1046/j.1365-8711.2003.05925.x}, 338, 14

\bibitem[\protect\citeauthoryear{Prada, Klypin, Cuesta, Betancort-Rijo  \&
  Primack}{Prada et~al.}{2012}]{Prada2012}
Prada F.,  Klypin A.~A.,  Cuesta A.~J.,  Betancort-Rijo J.~E.,   Primack J.,
  2012, \mn@doi [\mnras] {10.1111/j.1365-2966.2012.21007.x}, 423, 3018

\bibitem[\protect\citeauthoryear{Price}{Price}{2012}]{Price2012}
Price D.~J.,  2012, \mn@doi [Journal of Computational Physics]
  {10.1016/j.jcp.2010.12.011}, 231, 759

\bibitem[\protect\citeauthoryear{Price et~al.,}{Price et~al.}{2018}]{Price2018}
Price D.~J.,  et~al., 2018, \mn@doi [Publications of the Astronomical Society
  of Australia] {10.1017/pasa.2018.25}, 35

\bibitem[\protect\citeauthoryear{Read, Hayfield  \& Agertz}{Read
  et~al.}{2010}]{Read2010}
Read J.~I.,  Hayfield T.,   Agertz O.,  2010, \mn@doi [\mnras]
  {10.1111/j.1365-2966.2010.16577.x}, 405, 1513

\bibitem[\protect\citeauthoryear{Richardson, Quinn, Stadel  \& Lake}{Richardson
  et~al.}{2000}]{Richardson2000}
Richardson D.,  Quinn T.,  Stadel J.,   Lake G.,  2000, \mn@doi [\icarus]
  {10.1006/icar.1999.6243}, 143, 45

\bibitem[\protect\citeauthoryear{Rosito et~al.,}{Rosito
  et~al.}{2021}]{Rosito2020}
Rosito M.~S.,  et~al., 2021, \mn@doi [\aap] {10.1051/0004-6361/202039976}, 652,
  A44

\bibitem[\protect\citeauthoryear{Rosswog}{Rosswog}{2009}]{Rosswog2009}
Rosswog S.,  2009, \mn@doi [Advances in the Free-Lagrange Method Including
  Contributions on Adaptive Gridding and the Smooth Particle Hydrodynamics
  Method] {10.1016/j.newar.2009.08.007}, pp 239--247

\bibitem[\protect\citeauthoryear{Ryu, Ostriker, Kang  \& Cen}{Ryu
  et~al.}{1993}]{Ryu1993}
Ryu D.,  Ostriker J.~P.,  Kang H.,   Cen R.,  1993, \mn@doi [\apj]
  {10.1086/173051}, 414, 1

\bibitem[\protect\citeauthoryear{Saitoh \& Makino}{Saitoh \&
  Makino}{2009}]{Saitoh09}
Saitoh T.~R.,  Makino J.,  2009, \mn@doi [\apj] {10.1088/0004-637X/697/2/L99},
  697, 99

\bibitem[\protect\citeauthoryear{Schaller, Gonnet, Chalk  \& Draper}{Schaller
  et~al.}{2016}]{Schaller2016}
Schaller M.,  Gonnet P.,  Chalk A.~B.,   Draper P.~W.,  2016, \mn@doi [PASC
  2016 - Proceedings of the Platform for Advanced Scientific Computing
  Conference] {10.1145/2929908.2929916}

\bibitem[\protect\citeauthoryear{Schaller, Gonnet, Draper, Chalk, Bower, Willis
   \& Hausammann}{Schaller et~al.}{2018}]{Schaller2018}
Schaller M.,  Gonnet P.,  Draper P.~W.,  Chalk A.~B.,  Bower R.~G.,  Willis J.,
    Hausammann L.,  2018, {SWIFT: SPH With Inter-dependent Fine-grained
  Tasking}, \url {https://ascl.net/1805.020}

\bibitem[\protect\citeauthoryear{Schaye et~al.,}{Schaye
  et~al.}{2010}]{Schaye2010}
Schaye J.,  et~al., 2010, \mn@doi [\mnras] {10.1111/j.1365-2966.2009.16029.x},
  402, 1536

\bibitem[\protect\citeauthoryear{Schaye et~al.,}{Schaye
  et~al.}{2015}]{Schaye2015}
Schaye J.,  et~al., 2015, \mn@doi [\mnras] {10.1093/mnras/stu2058}, 446, 521

\bibitem[\protect\citeauthoryear{Schneider et~al.,}{Schneider
  et~al.}{2016}]{Schneider2016}
Schneider A.,  et~al., 2016, \mn@doi [Journal of Cosmology and Astroparticle
  Physics] {10.1088/1475-7516/2016/04/047}, 2016, 047

\bibitem[\protect\citeauthoryear{Sedov}{Sedov}{1959}]{Sedov59}
Sedov L.~I.,  1959, {Similarity and Dimensional Methods in Mechanics}.
Elsevier, \mn@doi{10.1016/C2013-0-08173-X}, \url
  {https://linkinghub.elsevier.com/retrieve/pii/C2013008173X}

\bibitem[\protect\citeauthoryear{Sembolini et~al.,}{Sembolini
  et~al.}{2016a}]{Sembolini2016}
Sembolini F.,  et~al., 2016a, \mn@doi [\mnras] {10.1093/mnras/stw250}, 457,
  4063

\bibitem[\protect\citeauthoryear{Sembolini et~al.,}{Sembolini
  et~al.}{2016b}]{Sembolini2016b}
Sembolini F.,  et~al., 2016b, \mn@doi [\mnras] {10.1093/mnras/stw800}, 459,
  2973

\bibitem[\protect\citeauthoryear{Somerville \& Dav{\'{e}}}{Somerville \&
  Dav{\'{e}}}{2015}]{Somerville2015}
Somerville R.~S.,  Dav{\'{e}} R.,  2015, \mn@doi [\araa]
  {10.1146/annurev-astro-082812-140951}, 53, 51

\bibitem[\protect\citeauthoryear{Springel}{Springel}{2005}]{Springel2005b}
Springel V.,  2005, \mn@doi [\mnras] {10.1111/j.1365-2966.2005.09655.x}, 364,
  1105

\bibitem[\protect\citeauthoryear{Springel}{Springel}{2010}]{Springel2010}
Springel V.,  2010, \mn@doi [\mnras] {10.1111/j.1365-2966.2009.15715.x}, 401,
  791

\bibitem[\protect\citeauthoryear{Springel, Pakmor, Zier  \& Reinecke}{Springel
  et~al.}{2021}]{Springel2021}
Springel V.,  Pakmor R.,  Zier O.,   Reinecke M.,  2021, \mn@doi [\mnras]
  {10.1093/mnras/stab1855}, 506, 2871

\bibitem[\protect\citeauthoryear{Stadel}{Stadel}{2001}]{Stadel2001}
Stadel J.,  2001, PhD thesis, University of Washington, \url
  {http://adsabs.harvard.edu/abs/2001PhDT........21S}

\bibitem[\protect\citeauthoryear{Steinmetz \& M{\"{u}}ller}{Steinmetz \&
  M{\"{u}}ller}{1993}]{Steinmetz1993}
Steinmetz M.,  M{\"{u}}ller E.,  1993, \aap, 268, 391

\bibitem[\protect\citeauthoryear{Steinmetz \& White}{Steinmetz \&
  White}{1997}]{Steinmetz1997}
Steinmetz M.,  White S.~D.,  1997, \mn@doi [\mnras] {10.1093/mnras/288.3.545},
  288, 545

\bibitem[\protect\citeauthoryear{Stone, Gardiner, Teuben, Hawley  \&
  Simon}{Stone et~al.}{2008}]{Stone2008}
Stone J.~M.,  Gardiner T.~A.,  Teuben P.,  Hawley J.~F.,   Simon J.~B.,  2008,
  \mn@doi [\apjs] {10.1086/588755}, 178, 137

\bibitem[\protect\citeauthoryear{Teyssier}{Teyssier}{2002}]{Teyssier2002}
Teyssier R.,  2002, \mn@doi [\aap] {10.1051/0004-6361:20011817}, 385, 337

\bibitem[\protect\citeauthoryear{{The HDF Group}}{{The HDF Group}}{2022}]{hdf5}
{The HDF Group} 2022, {Hierarchical Data Format, version 5}

\bibitem[\protect\citeauthoryear{Toro}{Toro}{2009}]{Toro2009}
Toro E.~F.,  2009, {Riemann Solvers and Numerical Methods for Fluid Dynamics}.
Springer Berlin Heidelberg, Berlin, Heidelberg, \mn@doi{10.1007/b79761}, \url
  {http://link.springer.com/10.1007/b79761}

\bibitem[\protect\citeauthoryear{Vandenbroucke \& {De Rijcke}}{Vandenbroucke \&
  {De Rijcke}}{2016}]{Vandenbroucke2016}
Vandenbroucke B.,  {De Rijcke} S.,  2016, \mn@doi [Astronomy and Computing]
  {10.1016/j.ascom.2016.05.001}, 16, 109

\bibitem[\protect\citeauthoryear{Vila}{Vila}{1999}]{Vila99}
Vila J.~P.,  1999, \mn@doi [Mathematical Models and Methods in Applied
  Sciences] {10.1142/S0218202599000117}, 09, 161

\bibitem[\protect\citeauthoryear{Vogelsberger et~al.,}{Vogelsberger
  et~al.}{2014}]{Vogelsberger2014}
Vogelsberger M.,  et~al., 2014, \mn@doi [\mnras] {10.1093/mnras/stu1536}, 444,
  1518

\bibitem[\protect\citeauthoryear{Vogelsberger, Marinacci, Torrey  \&
  Puchwein}{Vogelsberger et~al.}{2020}]{Vogelsberger2019}
Vogelsberger M.,  Marinacci F.,  Torrey P.,   Puchwein E.,  2020, \mn@doi [\nat
  Reviews Physics] {10.1038/s42254-019-0127-2}, 2, 42

\bibitem[\protect\citeauthoryear{Wadsley, Stadel  \& Quinn}{Wadsley
  et~al.}{2004}]{Wadsley2004}
Wadsley J.,  Stadel J.,   Quinn T.,  2004, \mn@doi [\na]
  {10.1016/j.newast.2003.08.004}, 9, 137

\bibitem[\protect\citeauthoryear{Wadsley, Keller  \& Quinn}{Wadsley
  et~al.}{2017}]{Wadsley2017}
Wadsley J.~W.,  Keller B.~W.,   Quinn T.~R.,  2017, \mn@doi [\mnras]
  {10.1093/mnras/stx1643}, 471, 2357

\bibitem[\protect\citeauthoryear{Wang, Dutton, Stinson, Macci{\`{o}}, Penzo,
  Kang, Keller  \& Wadsley}{Wang et~al.}{2015}]{Wang2015}
Wang L.,  Dutton A.~A.,  Stinson G.~S.,  Macci{\`{o}} A.~V.,  Penzo C.,  Kang
  X.,  Keller B.~W.,   Wadsley J.,  2015, \mn@doi [\mnras]
  {10.1093/mnras/stv1937}, 454, 83

\bibitem[\protect\citeauthoryear{Weinberger, Springel  \& Pakmor}{Weinberger
  et~al.}{2020}]{Weinberger2020}
Weinberger R.,  Springel V.,   Pakmor R.,  2020, \mn@doi [\apjs]
  {10.3847/1538-4365/ab908c}, 248, 32

\bibitem[\protect\citeauthoryear{Zeldovich}{Zeldovich}{1970}]{Zeldovich1970}
Zeldovich Y.~B.,  1970, \aap, 5, 84

\makeatother
\end{thebibliography}

%%%%%%%%%%
%%%%%%%%%%
%%%%%%%%%%

\end{document}